\documentclass[twocolumn,floatfix]{revtex4-2}

\usepackage{amsmath}
\usepackage{amssymb}
\usepackage{amsfonts,amsthm,bm}
\usepackage{xfrac}
\usepackage{nicematrix}
\usepackage[caption=false]{subfig}
\usepackage{mathrsfs}
\usepackage{tikz}
\usepackage{pgfplots}
\pgfplotsset{compat=1.18}
\usetikzlibrary{arrows.meta}
\usetikzlibrary{external}
\usepackage{standalone}
\usepackage{hyperref}

\newcommand{\hamil}{\mathcal{H}}

\newcommand{\euler}[1]{\mathrm{e}^{#1}}

\newcommand{\timeorder}{\mathcal{T}}

\renewcommand{\vec}{\mathbf}

\newcommand{\freephon}{\mathfrak{D}}
\newcommand{\gloc}{\mathcal{G}}

\renewcommand{\vec}{\mathbf}
\newcommand{\gfree}{\mathfrak{G}}

\allowdisplaybreaks

\begin{document}

\begin{abstract}

The interplay between spin and orbital degrees of freedom gives rise to a variety of emergent phases in correlated 4$d$ and 5$d$ transition-metal systems. Strong spin--orbit coupling (SOC) significantly alters Jahn-Teller (JT) physics, often suppressing static distortions or promoting dynamic fluctuations, thereby reducing or even quenching orbital polarization. While intersite hybridization is a fundamental aspect of crystalline solids, its role in shaping the dynamics of spin-orbit-entangled states has received comparatively little attention. Here, we show that electronic hopping can locally restore orbital polarization when the ground state is perturbed, even in the absence of static orbital order. Using a Matsubara lattice formalism, we analyze how local orbital perturbations propagate through correlated, spin-orbit-entangled systems. When intersite hopping is included, such perturbations induce short-range orbital polarization with a characteristic orthogonal response at nearest-neighbor sites. Although the energy scale of these hybridization-driven orbital reconstructions likely makes their detection challenging, they may still influence low-energy spectral features and interact with other excitations. These results underscore the importance of including orbital dynamics in the interpretation of spectroscopic data and provide a framework for understanding dynamical responses in spin-orbit-entangled materials.

\end{abstract}

\author{A. S. Mi{\~n}arro, G. Herranz}
\title{Emergent Orbital Dynamics in Strongly Spin-Orbit Coupled Systems}
\affiliation{Institut de Ci\`encia de Materials de Barcelona (ICMAB-CSIC), Campus UAB, 08193 Bellaterra, Catalonia, Spain}
\maketitle

\begin{figure*}[t]
    \centering
    \includegraphics[width=0.7\textwidth]{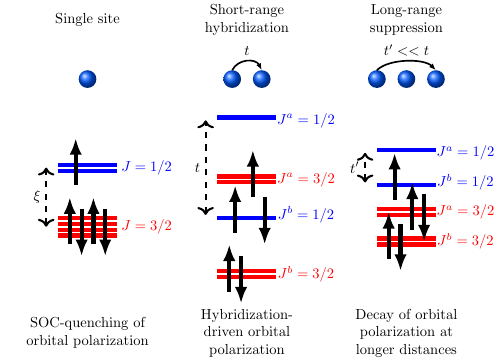}
\caption{
Schematic illustration of how finite hopping amplitude may reinstate local orbital polarization in the $t_{2g}$ shell with five electrons.  
(Left) In the atomic limit, strong spin--orbit coupling (SOC, \( \xi \)) lifts orbital degeneracy and quenches static Jahn--Teller (JT) distortions, placing the fifth electron in a spin--orbit entangled \( j = 1/2 \) state. The \( j=1/2 \) wavefunctions are fully entangled combinations of the $t_{2g}$ orbitals, with equal weights for \( |xy\rangle \), \( |yz\rangle \), and \( |zx\rangle \), preventing orbital polarization.  
(Middle) In the lattice, a local perturbation interacts with neighboring sites via finite hopping, inducing bonding (b) and antibonding (a) states. The antibonding electron acquires partial \( j=3/2 \) character, involving states such as \( |j=3/2, m_j=+3/2\rangle \sim (|yz, \uparrow\rangle + i |zx, \uparrow\rangle) \), which retain orbital anisotropy and allow hybridization to selectively polarize orbitals. This local orbital polarization arises from kinetic hybridization effects, even in the absence of explicit orbital--lattice coupling. The schematic focuses on the local orbital response induced by a perturbation, without attempting to represent the fully delocalized band structure.  
(Right) Beyond first neighbors, hybridization is too weak to overcome SOC splitting, and orbital polarization vanishes. As a result, the effect is inherently short-ranged, confined to a few lattice spacings.
}
\label{fig1sketch}
\end{figure*}

Strong spin-orbit coupling in $t_{2g}$ states provides fertile ground for exploring various emergent phases in transition metal compounds, including spin-orbit Mott insulators, spin liquids, multipolar orders, and excitonic magnetism \cite{rau2016spin,takayama2021spin,trebst2022kitaev}. The interplay between spin, lattice and orbital degrees of freedom, along with electronic correlations, leads to complex phase diagrams, particularly in 4$d$ and 5$d$ systems \cite{brzezicki2019spin,khomskii2020orbital,rousochatzakis2024beyond}. Understanding how orbital degeneracy is lifted is crucial for uncovering the nature of spin-orbit entangled states and their influence on physical properties \cite{gotfryd2020spin,iwahara2024dynamic}.

Although electronic correlations are widely recognized to shape spin-orbital entanglement~\cite{witczak2014correlated,takagi2019concept,schaffer2016recent,PhysRevResearch.3.033163,georges2013strong}, the role of orbital-lattice coupling, particularly Jahn-Teller (JT) interactions, in the presence of strong spin-orbit coupling (SOC) has traditionally received less attention. In the conventional view, SOC is thought to quench JT distortions by stabilizing spin-orbit entangled states that suppress orbital polarization. However, recent studies suggest a more nuanced scenario in which the effects of JT may persist in a dynamic form even under strong SOC~\cite{celiberti2024spin,vzivkovic2024dynamic}. This opens the possibility that SOC and orbital polarization tendencies may, in certain regimes, cooperate rather than compete~\cite{minarro2022jahn, minarro2024spin,vzivkovic2024dynamic,iwahara2024dynamic}.

On the other hand, while intersite electronic hybridization is an intrinsic feature of any crystalline lattice and is naturally incorporated in electronic structure models, its specific impact on the orbital dynamics of spin–orbit entangled states has not been directly addressed. Interestingly, our work demonstrates that short-range orbital polarization can dynamically reemerge in response to local perturbations, even when static orbital order is suppressed by strong SOC. This effect originates from interorbital hopping processes in multiorbital systems and does not rely on lattice distortions or explicit symmetry breaking.

To address these issues, we employ the Matsubara Green’s function formalism~\cite{mahan2000many,aoki2014nonequilibrium,abrikosov1963methods,fetter1971theory}, exploring parameter regimes relevant to correlated \( t_{2g} \) systems, including spin--orbit coupling, electronic correlations, and intersite hopping. We analyze how a local orbital perturbation propagates through the system and find that it induces short-range orbital correlations in an otherwise fully symmetric ground state, with polarization effects confined to nearest-neighbor sites. Although SOC typically suppresses static Jahn--Teller distortions, our results show that local perturbations can dynamically restore orbital anisotropy, revealing latent orbital structure embedded in the spin--orbit-entangled background (see Fig.~\ref{fig1sketch} for an illustrative case with five electrons in the \( t_{2g} \) shell). \\
\\
\noindent
While the energy scale of these fluctuations makes their direct experimental detection quite challenging, we discuss their potential influence on low-energy spectral features in correlated materials. Taken together, our results suggest that the absence of long-range orbital order does not necessarily preclude the presence of local orbital fluctuations, which may emerge dynamically through short-range hybridization processes in response to local perturbations. This points to the value of theoretical approaches that go beyond static mean-field treatments, incorporating the interplay between spin--orbit coupling, itinerancy, and latent orbital instabilities. The framework we present may complement existing methods, such as single-site and cluster dynamical mean-field theory (DMFT)~\cite{eckstein2021simulation, hariki2020lda+, murakami2023photo}, by providing additional insight into short-range orbital polarization effects that are difficult to capture explicitly in spin-orbit-coupled systems.

\begin{figure*}[ht]
    \centering
    \includegraphics[width=\textwidth]{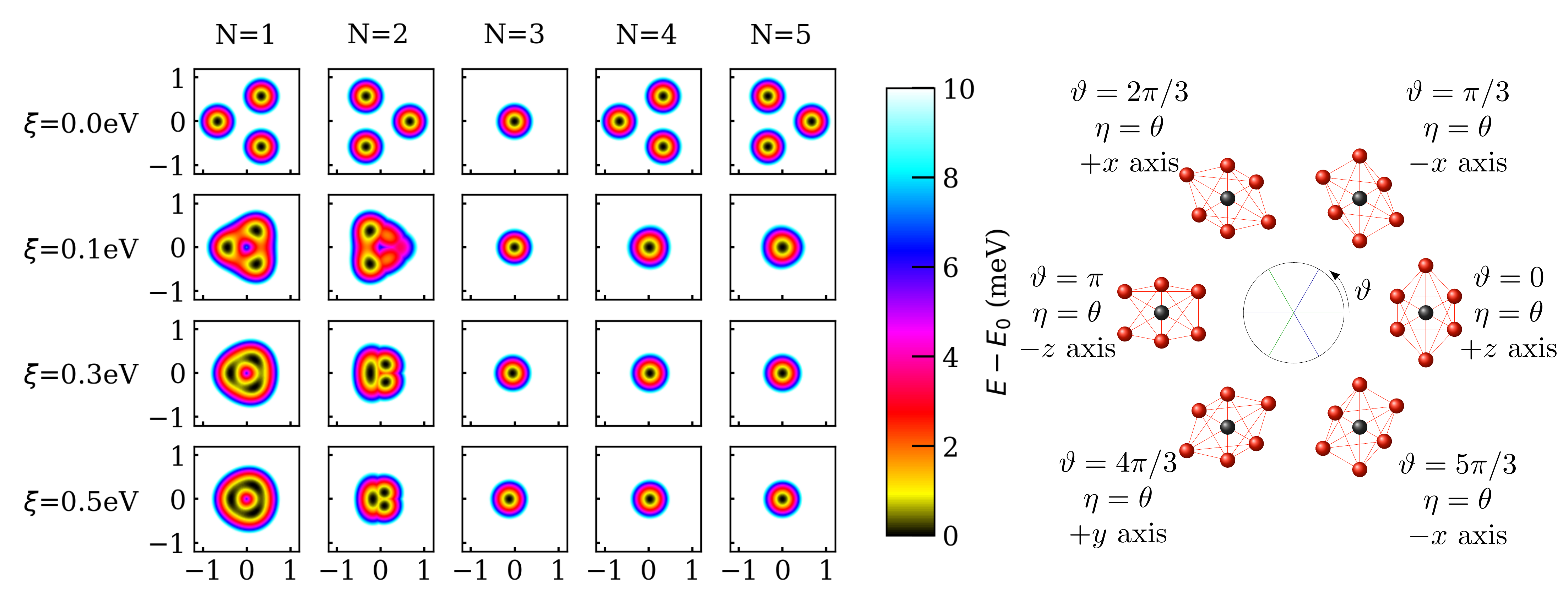}
  \caption{Single-site model: competition between spin-orbit coupling and Jahn-Teller distortions. Energy contours for different orbital occupancies $d^N$ as a function of spin-orbit coupling strength $\xi$, mapped in the Jahn-Teller distortion plane $(Q_3,Q_2)$. Pure tetragonal modes are represented by $\vartheta = 0, \pi, \pm\frac{\pi}{3}, \pm\frac{2\pi}{3}$, while pure orthorhombic modes by $\vartheta = \pm\frac{\pi}{2}, \pm\frac{\pi}{6}, \pm\frac{5\pi}{6}$. The sketches on the right illustrate the distortion patterns for elongated $(+)$ and compressed $(-)$ tetragonal distortions $(\eta = \theta)$ along the three orthogonal directions $\hat{x}, \hat{y}, \hat{z}$. The radial coordinate $Q$ represents the distortion amplitude, while the angular coordinate $\vartheta$ distinguishes contributions from tetragonal and orthorhombic modes of $E_g$ symmetry. The energy, encoded in the color bar, is truncated at 10~meV above the minimum to enhance contrast and better visualize the location of energy minima. }
\label{fig1}
\end{figure*}

\section{Single-site Hamiltonian} 
We first examine a single-site model that describes the Jahn-Teller coupling of \( t_{2g} \) electrons with \( E_g \) lattice modes (the \( t \otimes E \) problem). Within this single-site framework, the Jahn-Teller effects are described by \cite{kugel1982jahn, strestolvPhysRevX.10.031043}
\begin{align*}
    \begin{split}
	\mathcal{H}_{JT} &= -Qg \left\{\dfrac{1}{\sqrt{3}}\left[l_x^2 - l_y^2\right] \sin\vartheta + \left[ l_z^2 - \dfrac{2}{3} \right] \cos\vartheta \right\} \\ 
    &\phantom{=} + \dfrac{B}{2}Q^2
    \end{split}
\end{align*}
where $g,B$ are coupling constants, $l_\alpha$ are the angular momentum operators. The Jahn-Teller term is written in terms of polar coordinates $Q_3 = Q\cos\vartheta$ and $Q_2 = Q\sin\vartheta$ (Figure \ref{fig1}), where $Q_2,Q_3$ are, respectively, the orthorhombic and tetragonal coordinates of the $t \otimes E$ Jahn-Teller vibronic modes \cite{bersuker2013jahn, bersuker1982vibronic}. Electron - electron interactions are expressed through the Kanamori model 
\begin{align*}
\begin{split}
\mathcal{H}_{ee} &=
    U\sum_{\alpha} n_{\alpha\uparrow} n_{\alpha\downarrow} + U'\sum_{\alpha\neq\gamma} n_{\alpha\uparrow} n_{\gamma\downarrow} \\ 
    &\phantom{=} + (U'-J)\sum_{\alpha<\gamma,\sigma} n_{\alpha\sigma} n_{\gamma\sigma} \\ 
    &\phantom{=} - J\sum_{\alpha\neq\gamma} c_{\alpha\uparrow}^\dagger \left( c_{\alpha\downarrow} c_{\gamma\downarrow}^\dagger + c_{\gamma\downarrow} c_{\alpha\downarrow}^\dagger \right) c_{\gamma\uparrow}
\end{split}
\end{align*}
which includes intra-orbital \(U\), inter-orbital \(U'\) Coulomb repulsions, and the Hund's coupling \(J\). Here, \(\alpha\) and \(\gamma\) denote orbitals within the \( t_{2g} \) manifold, while \(c_{\alpha}, c_{\alpha}^\dagger\) and \(c_{\gamma}, c_{\gamma}^\dagger\) represent the annihilation and creation ladder operators, respectively. The number operator is defined as $n_\alpha = c_{\alpha\uparrow}^\dagger c_{\alpha\uparrow} + c_{\alpha\downarrow}^\dagger c_{\alpha\downarrow}$. Spin-orbit coupling is included through a term
\[
\hamil_{SO} = \frac{\xi}{2} \sum_i \sum_{\sigma,\sigma'} \sum_{\mu,\nu \in d} 
        \langle \mu|\vec{l}|\nu\rangle \cdot \langle \sigma|\boldsymbol{\sigma}|\sigma'\rangle 
     c_{i\mu\sigma}^\dagger c_{i\nu\sigma'}
\]
where $\xi$ is the SOC strength, $\vec{l}$ the orbital angular momentum and $\boldsymbol{\sigma}$ the spin. The sums extend over lattice sites $i$, spin states $\sigma, \sigma'$, and orbitals $\mu, \nu$ within the $t_{2g}$ manifold.  In our calculations we used the values $\xi = 0.1 - 0.5$ eV, $U = 2.5$ eV, $J = 0.2$ \cite{li2013atomically, ma2014systematic}.

We first discuss the case of single electron occupancy (\( N = 1 \)). The spin-orbit coupling is given by $\hamil_{SOC} = -\dfrac{\xi}{2} \vec{l} \cdot \boldsymbol{\sigma}$, where \( \xi \) is the SOC constant, and $\boldsymbol{\sigma}$ represents the Pauli vector. SOC splits the \( t_{2g} \) multiplet into a higher-energy \( j = 1/2 \) doublet and a lower-energy \( j = 3/2 \) quartet. For strong SOC (\( \xi \gg g^2/B \)), the JT interaction has no first-order effect on \( j = 1/2 \) states. For \( j = 3/2 \), first-order perturbation theory reveals eigenvalue subspaces with an excited state at \( Qg/3 \) and a ground state at \( -Qg/3 \), which maps to the \( e \otimes E \) problem associated with a "Mexican hat" energy landscape \cite{bersuker2012vibronic, bersuker2013jahn, khomskii2014transition}. Our calculations confirm this point, illustrated by the energy contours in polar coordinates \( (Q, \vartheta) \) for \( g = B = 0.1 \, \text{eV} \) (Figure \ref{fig1}). For relatively weak SOC ($\xi \leq 0.3$ eV), three minima appear at \( \vartheta = \pi, \frac{5\pi}{3}, \frac{\pi}{3} \), corresponding to compressed JT distortions. However, with increasing SOC, these minima gradually vanish, and the energy landscape tends to a ring-shaped profile ~\cite{strestolvPhysRevX.10.031043}. This potential is compatible with dynamic JT effects \cite{iwahara2023vibronic}, which have been claimed in \( d^1 \) systems (e.g., Re$^{6+}$, Os$^{7+}$), particularly in resonant inelastic X-ray scattering (RIXS) experiments \cite{iwahara2024dynamic, iwahara2024persistent, frontini2024spin, agrestiniPhysRevLett.133.066501, vzivkovic2024dynamic}.

For \(N = 2\) and moderate SOC (\(\xi = 0.1 \, \text{eV}\)), minima correspond to elongated tetragonal distortions \cite{strestolvPhysRevX.10.031043}. As SOC increases (\(\xi = 0.3 - 0.5 \, \text{eV}\)), compressed distortions dominate. Therefore, for realistic values of SOC, JT distortions are not fully suppressed in this case (Figure \ref{fig1}). Nevertheless, quantum tunneling between local minima may enable dynamical Jahn-Teller effects. In fact, evidence from RIXS spectra in systems like Ba$_2$YReO$_6$ and Ba$_2$CaOsO$_6$ suggests the possibility of such effects \cite{iwahara2024dynamic}, and the competition between spin-orbit coupling, multipolar exchange, and Jahn-Teller effects may promote dynamically fluctuating distortions characteristic of dynamical Jahn-Teller effects \cite{khaliullin2021exchange}.

For three electrons (\(N = 3\)) in the \(t_{2g}\) shell, the \(j = 3/2\) levels dominate under strong SOC. For typical SOC strengths (\(\xi \approx 0.1 - 0.3 \, \text{eV}\)) and vibronic parameters (\(g = B = 0.1 \, \text{eV}\)), JT static distortions remain negligible (Figure \ref{fig1}). This finding is consistent with experimental studies on K$_2$ReCl$_6$ \cite{warzanowski2024spin}, which report no detectable JT effects at SOC strength $\xi \approx$ 0.29 eV, and is also supported by our calculations at similar SOC values. However, systems with significantly stronger SOC present a different scenario. Our calculations shown in Figure \ref{fig1} reveal that at higher SOC strengths (\(\xi \approx 0.5 \, \text{eV}\)), there is a meaningful JT distortion, aligning with theoretical predictions from Ref. \cite{strestolvPhysRevX.10.031043}. Therefore, to experimentally observe SOC-driven JT effects, candidate systems should possess notably larger SOC strength.

For four electrons (\(N = 4\)), the ground state (\(J = 0\)) in the strong SOC limit suppresses all JT distortions \cite{iwahara2024dynamic}. Calculations confirm that even for moderate SOC (\(\xi = 0.1 \, \text{eV}\)), JT distortions remain inactive (Figure \ref{fig1}), although dynamic JT effects may occur in excited \(J = 1, 2\) states or with \(e_g\)-state mixing \cite{iwahara2024dynamic, voleti2020multipolar, takayama2021spin}. For five electrons (\(N = 5\)), the \(J = 1/2\) ground state remains inactive to Jahn-Teller distortions, and this explains why spin-orbit coupling quickly suppresses them, in agreement with Ref.~\cite{strestolvPhysRevX.10.031043}. However, dynamic JT effects may still emerge in the excited \(J = 3/2\) states, which could be probed through techniques such as RIXS or Raman spectroscopy in compounds like K$_2$IrCl$_6$~\cite{iwahara2023vibronic, lee2022noncubic, iwahara2024dynamic}.

Consequently, the results of the single-site model confirm that for typical SOC strengths in $4d$ and $5d$ ions, JT effects are generally suppressed or, at best, persist as dynamic distortions, consistent with recent advanced spectroscopic experiments and implying a reduction or a quenching of orbital polarization. However, a comprehensive understanding of spin–orbital excitations must account for the effects of intersite electronic hybridization, which is an inherent feature of crystalline solids. In the following, we examine this important aspect.\\

\begin{figure*}[t] 
    \centering
    \includegraphics[width=\textwidth]{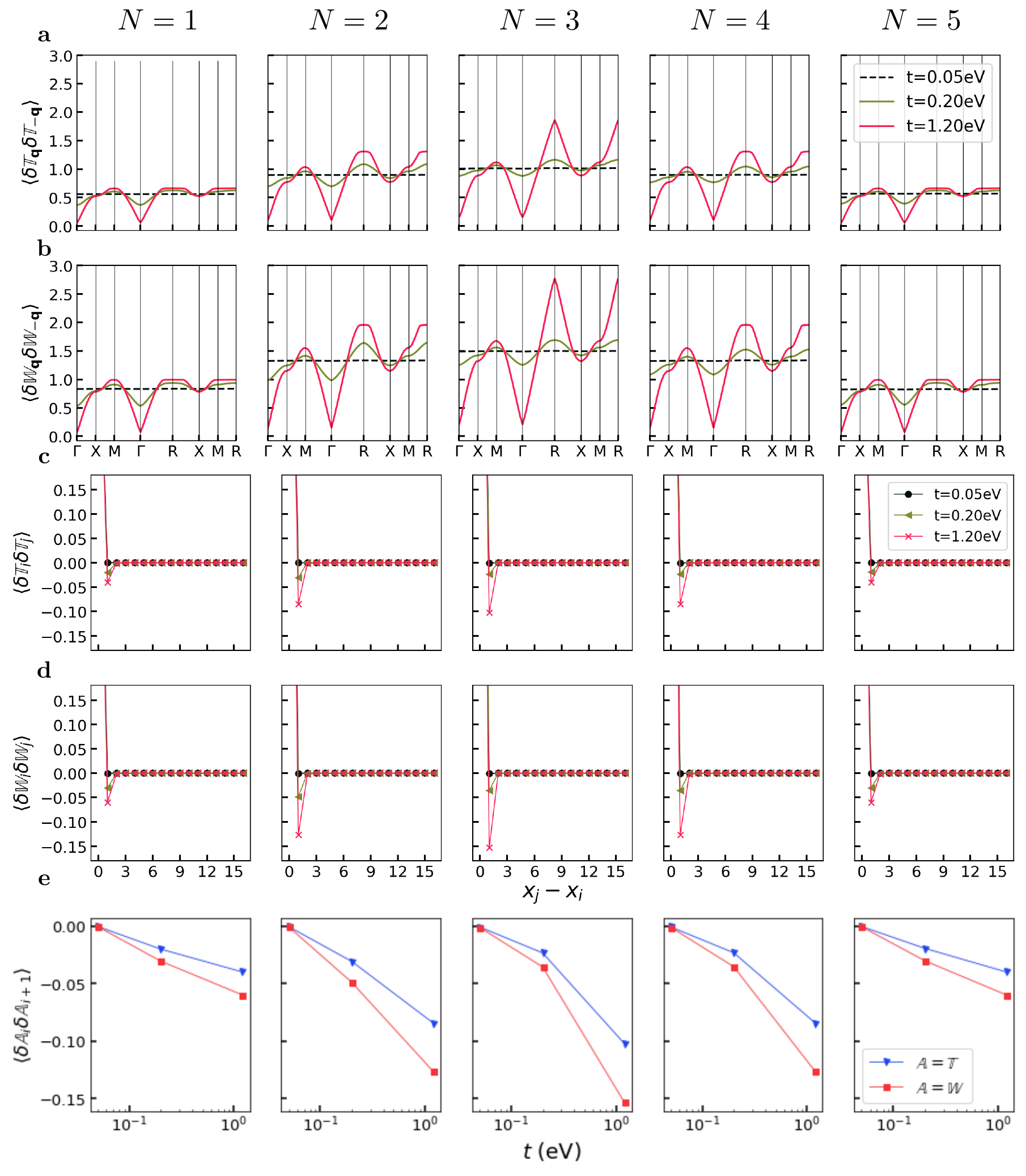}
    % \includegraphics[width=\textwidth]{Fig2a_4d.png} \\
    % \includegraphics[width=\textwidth]{Fig2b_4d.png} \\
    % \includegraphics[width=\textwidth]{Fig2c_4d.png}
    % \begin{tikzpicture}
    %     \node at (0,0) {\includegraphics[width=\textwidth]{Fig2a_4d.png}};
    %     \node at (0,-1) {\includegraphics[width=\textwidth]{Fig2b_4d.png}};
    %     \node at (0,-2) {\includegraphics[width=\textwidth]{Fig2c_4d.png}};
    % \end{tikzpicture}
    \caption{Spin-orbital correlations in the lattice model as a function of electron filling.  
(a) Momentum-space orbital-orbital correlations, $\langle \delta \mathbb{T}^\alpha_\vec{q} \delta \mathbb{T}^\alpha_\vec{-q} \rangle$, evaluated along high-symmetry points ($\Gamma, X, M, R$) of the Brillouin zone for different hopping amplitudes. (b) Momentum-space spin-orbital correlations, $\langle \delta \mathbb{W}_\vec{q} \delta \mathbb{W}_\vec{-q} \rangle$, computed under the same conditions as in (a). (c) Real-space orbital-orbital correlations, $\langle \delta \mathbb{T}_i \delta \mathbb{T}_j \rangle$, as a function of intersite distance $|x_j - x_i|$, confirming their short-range nature, primarily confined to nearest neighbors. The negative values indicate cooperative orbital alignment, where orbitals at nearest neighbors orient orthogonally to the one at the local perturbation site (Appendix \ref{app_corr}).  
(d) Real-space spin-orbital correlations, $\langle \delta \mathbb{W}_i \delta \mathbb{W}_j \rangle$, exhibiting a similar short-range behavior.  
(e) Nearest-neighbor spin-orbital ($A = \mathbb{W}$) and orbital-orbital ($A = \mathbb{T}$) correlations as a function of hopping amplitude \( t \), illustrating the suppression of short-range cooperative correlations as band hybridization weakens (lower \( t \)).  
All calculations were performed at \( T = 10\,\mathrm{K} \), corresponding to an inverse temperature \( \beta = 1 / k_{\mathrm{B}} T \approx 1160\,\mathrm{eV}^{-1} \), using parameters representative of 4d systems: \( g = B = 0.1\,\mathrm{eV} \), \( U = 2.5\,\mathrm{eV} \), and \( J = 0.4\,\mathrm{eV} \).}

    \label{Fig2M}
\end{figure*}

\section{Lattice model in Matsubara formalism}
 
To investigate the effects of spin-orbit and Jahn-Teller interactions in the lattice, we use a Matsubara Green’s function formalism. We consider the Hamiltonian
\begin{align*}
    \hamil &= \hamil_{latt} + \hamil_{JT} + \hamil_{ph} + \hamil_{SO} + \hamil_{ee} + \hamil_{oex} \\
    \hamil_{latt} &= \sum_{i,j} \sum_{\mu,\nu} \sum_\sigma t_{i\mu,j\nu} c_{i\alpha\sigma}^\dagger c_{j\beta\sigma} \\
    &= \sum_{\vec{k}} \sum_{\mu,\nu} \sum_\sigma \varepsilon_{\mu\nu}(\vec{k}) c_{\vec{k}\alpha\sigma}^\dagger c_{\vec{k}\beta\sigma} \\
    \hamil_{ph} &= \omega_0 \sum_i \left( b_{i\phi}^\dagger b_{i\phi} + b_{i\theta}^\dagger b_{i\theta} \right) \\
    \hamil_{JT} &= \sum_{i,\eta \in \{3,8\},\alpha,\beta,\sigma,\sigma'} g_\eta \lambda^\eta_{\alpha\beta} c_{i\alpha\sigma}^\dagger c_{i\beta\sigma'} \left( b_{i\eta} + b_{i\eta}^\dagger \right) \\
    &\quad \times \left( b_{i\eta} + b_{i\eta}^\dagger \right) \\
    \hamil_{SO} &= \frac{\xi}{2} \sum_i \sum_{\sigma,\sigma'} \sum_{\mu,\nu \in d} 
        \langle \mu|\vec{l}|\nu\rangle \cdot \langle \sigma|\boldsymbol{\sigma}|\sigma'\rangle \\
    &\quad \times c_{i\mu\sigma}^\dagger c_{i\nu\sigma'} \\
    \hamil_{ee} &= U \sum_{i,\alpha} n_{i\alpha\uparrow} n_{i\alpha\downarrow} 
        + U' \sum_{i,\alpha \neq \gamma} n_{i\alpha\uparrow} n_{i\gamma\downarrow} \\
    &\quad + (U' - J) \sum_{i,\alpha < \gamma,\sigma} n_{i\alpha\sigma} n_{i\gamma\sigma} \\
    &\quad - J \sum_{i,\alpha \neq \gamma} c_{i\alpha\uparrow}^\dagger 
        \left( c_{i\alpha\downarrow} c_{i\gamma\downarrow}^\dagger 
        + c_{i\gamma\downarrow} c_{i\alpha\downarrow}^\dagger \right) c_{i\gamma\uparrow} \\
    \hamil_{oex} &= - \sum_{i,j} \mathcal{J}_{ij} \sum_{\eta \in \{3,8\}} \sum_{\mu,\nu} 
        \lambda^\eta_{\mu\mu} \lambda^\eta_{\nu\nu} \sum_{\sigma,\sigma'} n_{i\mu\sigma} n_{j\nu\sigma'}
\end{align*}

We assume a cubic lattice with $t_{2g}$ electrons, incorporating a kinetic term ($\mathcal{H}_{\text{latt}}$), spin-orbit coupling ($\mathcal{H}_{\text{SO}}$), and electron-electron interactions ($\mathcal{H}_{\text{ee}}$). The indices $\mu\nu$ designate $t_{2g}$ orbitals, while $(i,j)$ are sites in the lattice. Furthermore, $\mathcal{H}_{\text{ph}}$ and $\mathcal{H}_{\text{JT}}$ describe the energy of the $E_g$ Jahn-Teller modes (with frequency $\omega_0$) and the $t \otimes E$ coupling (parameterized by $g_\eta$) between $t_{2g}$ electrons and the orthorhombic ($\eta = \phi$, $Q_2$) and tetragonal ($\eta = \theta$, $Q_3$) vibrational modes~\cite{strestolvPhysRevX.10.031043}. The last term, $ \hamil_{oex}$, accounts for orbital exchange arising from Jahn-Teller interactions mediated by the lattice. These interactions couple local Jahn-Teller distortions across multiple sites, resulting in an effective interorbital exchange \cite{khomskii2020orbital}. Both $\mathcal{H}_{\text{JT}}$ and $\mathcal{H}_{\text{oex}}$ involve the Gell-Mann matrices $\lambda^\eta_{\alpha\beta}$, which characterize the two $t \otimes E$ modes: $\eta = 3$ for the orthorhombic ($\phi$, $Q_2$) mode and $\eta = 8$ for the tetragonal ($\theta$, $Q_3$) mode. 

To solve the Hamiltonian, we evaluate the following self-consistent equations iteratively
\begin{align}
    \begin{split} \label{eq:GFself}
        \hat{G}(\imath\omega_n) &= \left[
        \hat{{\mathfrak{G}}}^{-1}(\imath\omega_n) - \hat{\Sigma}^{HF} 
        + \hat{\Sigma}^{(2)}(\imath\omega_n) \right. \\ 
        &\phantom{=} \left.
        + \hat{\Sigma}^{ep}(\imath\omega_n) + \hat{\Sigma}^{oex}(\imath\omega_n)
        \right]^{-1}
    \end{split} \\ 
    \label{eq:phself}
    D_\eta(\imath\nu_n) &=
    \left[
    \freephon^{-1}_\eta(\imath\nu_n) - \Pi_\eta(\imath\nu_n)
    \right]^{-1}
\end{align}
where the matrices $\hat{G}(\imath\omega_n)$ and $\hat{{\mathfrak{G}}}^{-1}(\imath\omega_n)$ are the dressed and bare electron Green's functions, respectively. The terms $\hat{\Sigma}^{HF}$ and $\hat{\Sigma}^{(2)}$ are the Hartree-Fock and second-order Born electronic self-energy matrices, while $\hat{\Sigma}^{ep}(\imath\omega_n)$ and $\hat{\Sigma}^{oex}(\imath\omega_n)$ are the electron-phonon and orbital exchange self-energies. Fermionic and bosonic frequencies are denoted by $\imath\omega_n$ and $\imath\nu_n$, respectively. Finally, $\freephon^{-1}_\eta(\imath\nu_n)$ represents the free phonon propagator, $D_\eta(\imath\nu_n)$ the dressed phonon propagator and $\Pi_\eta(\imath\nu_n)$ the phonon self-energy. Explicit expressions for the self-energies are found in Appendix \ref{MatInv}.\\

\begin{figure*}[tb] 
    \centering
    \includegraphics[width=\textwidth]{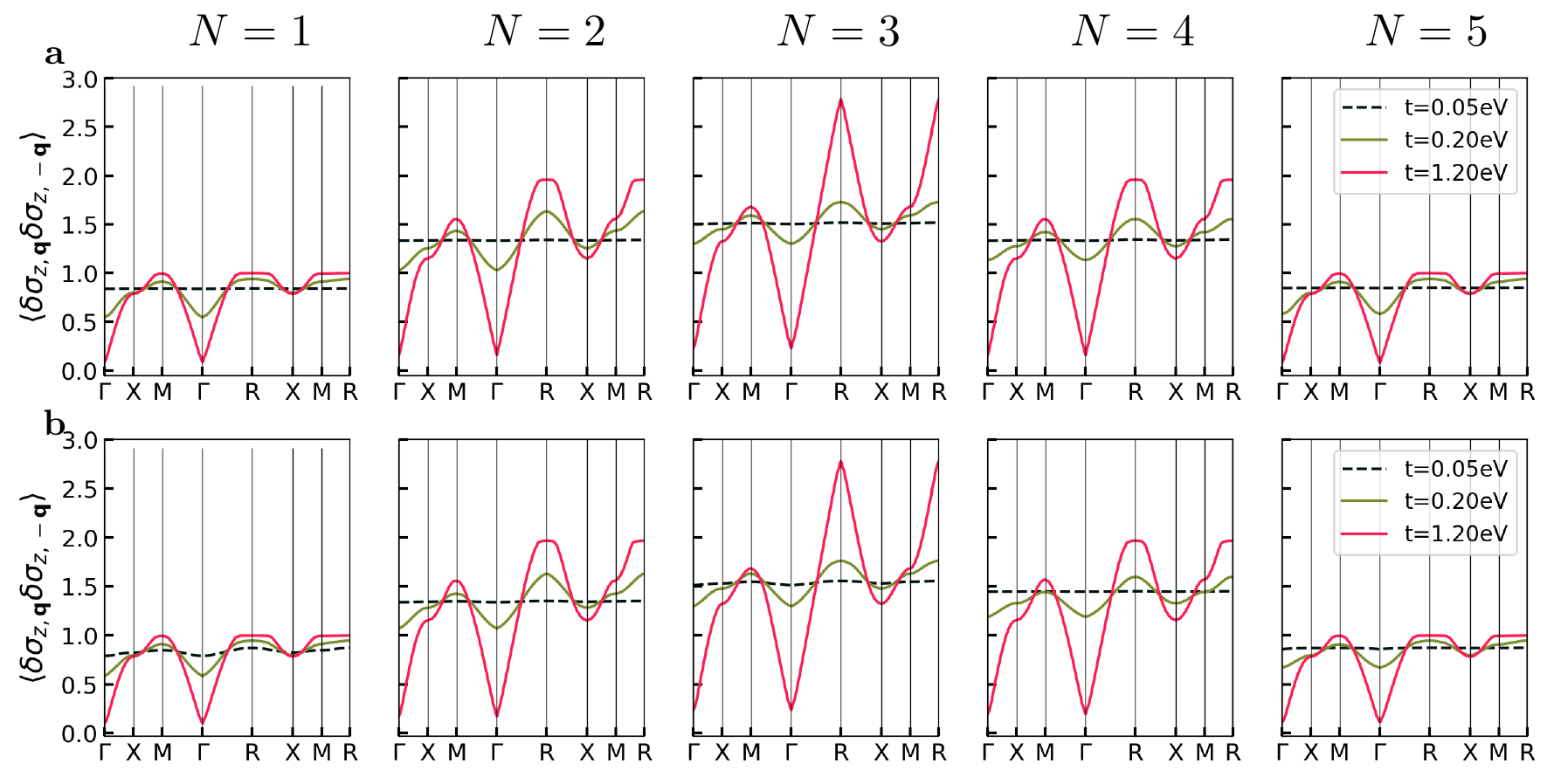}
\caption{
Momentum-resolved spin--spin correlations $\langle \delta \sigma^z_{\vec{q}} \, \delta \sigma^z_{-\vec{q}} \rangle$ for orbital fillings $N = 1$ to $5$. 
(a) Parameters representative of 4$d$ systems: $U = 2.5$ eV, $J = 0.4$ eV, spin--orbit coupling $\xi = 0.1$ eV, and Jahn--Teller coupling $g = 0.1$ eV. 
(b) Parameters typical of 5$d$ systems: $U = 2$ eV, $J = 0.2$ eV, $\xi = 0.3$ eV, and $g = 0.01$ eV. 
Each panel displays results for three hopping amplitudes, $t = 1.2$ eV (red), $t = 0.2$ eV (green), and $t = 0.05$ eV (black, dashed). 
The smallest value for the hopping amplitude, $t = 0.05$ eV, is not representative of real materials but illustrates how diminished hybridization suppresses spin coherence. 
As hopping increases, intersite exchange is enhanced, leading to the emergence of short-range antiferromagnetic correlations, reflected in prominent peaks at high-symmetry points such as $M = (\pi, \pi, 0)$, $R = (\pi, \pi, \pi)$, and $X = (\pi, 0, 0)$.
}

    \label{fig3M} % Replace 'figcorrel' with a meaningful label
\end{figure*}

\section{Orbital, spin-orbital and spin-spin correlations}
Spin--orbital correlations are computed using the orbital charge moment operator~\cite{kaga2022eliashberg}
\[
\mathbb{T}_{i\eta}(n) = \sum_{\alpha,\alpha',\sigma} \lambda_{\alpha\alpha'}^\eta(n)\, c_{i\alpha\sigma}^\dagger c_{i\alpha'\sigma},
\]
Here, $\hat{\lambda}^\eta(n)$ is constructed as a linear combination of the Gell-Mann matrices $\hat{\lambda}^3$ and $\hat{\lambda}^8$, transforming according to the irreducible representation $E_g$ (Appendix~\ref{app_corr}). The index $\eta = \phi, \theta$ labels the Jahn-Teller modes, and $n = 0,1,2$ refers to spatial directions $(x, y, z)$. By defining orbital charge moments in this form, we probe \( E_g \)-like orbital fluctuations in both real and momentum space. The correlators \( \langle \delta \mathbb{T} \, \delta \mathbb{T} \rangle \), with \( \delta \mathbb{T} = \mathbb{T} - \langle \mathbb{T} \rangle \), measure how the ground state responds to a local orbital perturbation. We note that the ground state is computed without introducing any explicit orbital symmetry breaking, i.e., it averages over all possible orientations of JT distortions when such distortions are energetically allowed (e.g., at filling \( N = 2 \), where nonzero JT distortions may arise for the SOC and JT coupling values used in Fig.~\ref{fig1}). As a result, the ground state remains rotationally symmetric, with \( \langle \mathbb{T} \rangle = 0 \), and our correlators capture the intrinsic orbital response without bias toward any specific polarization direction.

Orbital--orbital correlations in momentum space are expressed as (Appendix~\ref{app_corr})
\begin{align}
    \left\langle \delta \mathbb{T}_{\vec{q}\eta}(m)\, \delta \mathbb{T}_{-\vec{q}\zeta}(n) \right\rangle 
    \approx \dfrac{1}{\mathcal{N}}\sum_{\vec{k}} \mathrm{Tr} \big[ 
    \hat{\lambda}^\eta(m)\, \hat{G}_{\vec{k}}(0)\, \nonumber \\
    \times\, \hat{\lambda}^\zeta(n)\, \hat{G}_{\vec{k}-\vec{q}}(\beta) 
    \big],
\end{align}

where $\vec{k}, \vec{q}$ are momentum vectors. As shown in Fig.~\ref{Fig2M}a, the orbital-orbital correlators are progressively suppressed at the $\Gamma$ point with increasing hopping amplitude. We considered $t = 0.2$~eV and $t = 1.2$~eV, values within the typical range for 4$d$ and 5$d$ compounds~\cite{streltsov2017suppression, hafez2024antiferromagnetic}. This behavior reflects a significant reduction in long-range orbital correlations in both cases. However, pronounced maxima appear around high-symmetry points $R = (\pi/a, \pi/a, \pi/a)$ and $M = (\pi/a, \pi/a, 0)$, with the $R$-point particularly enhanced for fillings $N = 2$, 3, and 4. This momentum-space structure suggests a strong tendency toward staggered short-range orbital alignments, consistent with cooperative orbital responses to local perturbations, an interpretation further supported by the real-space correlation analysis discussed below. Consistently, the computed spin-orbital correlators $\langle \delta \mathbb{W}_{\vec{q}i}\, \delta \mathbb{W}_{-\vec{q}j} \rangle$ (Fig.~\ref{Fig2M}b) mirror the momentum-space structure of the orbital correlations, highlighting the intrinsic coupling between orbital and spin-orbital fluctuations~\cite{oles2003orbital, feiner2005orbital}. In particular, both sets of correlators exhibit robust short-range character, underscoring the interplay between these degrees of freedom.

The real-space orbital--orbital correlators $\langle \delta \mathbb{T}_i \, \delta \mathbb{T}_j \rangle$ (Figure~\ref{Fig2M}c) and spin--orbital correlators $\langle \delta \mathbb{W}_i \, \delta \mathbb{W}_j \rangle$ (Figure~\ref{Fig2M}d) offer further insight into the emergence of local orbital polarization. Their nonzero values confirm a short-range response to a localized orbital perturbation. The positive peak at $x = 0$ reflects the trivial autocorrelation of the perturbation itself, while the negative values at first neighbors reveal a moderate yet robust cooperative orbital alignment in which nearby orbitals tend to orient orthogonally (see Appendix~\ref{app_corr} for details on the interpretation of negative $\langle \delta \mathbb{T}_i \, \delta \mathbb{T}_j \rangle$). 

Remarkably, these orthogonal correlations persist even when spin-orbit coupling is strong ($\xi \geq 0.1 \, \text{eV}$) and static Jahn-Teller distortions are fully quenched in the atomic limit for $N = 3, 4, 5$ (see Figure~\ref{fig1}). This behavior suggests that the orbital polarization is not a consequence of orbital-lattice coupling, but rather emerges as an electronic effect, reinstated by intersite band hybridization. Even in systems representative of 5$d$ materials, where Jahn-Teller coupling is weak ($g = 0.01 \, \text{eV}$), similar short-range orbital correlations are observed (data not shown). These results point to a latent orbital instability, formally analogous to Jahn-Teller physics, that arises from hybridization-driven effects, even in the limit $g \to 0$. Additional information is obtained by examining the weak hopping regime ($t = 0.05 \, \text{eV}$, Figure~\ref{Fig2M}e), which lies below realistic values for most 4$d$ and 5$d$ systems. In this limit, the suppression of hybridization leads to the gradual disappearance of orbital polarization, smoothly recovering the fully symmetric limit of the ground state with vanishing orbital polarization. This provides compelling evidence that intersite electronic hybridization, enabled by finite hopping amplitudes, is the driving mechanism behind the emergence of short-range orbital polarization in the correlated \( t_{2g} \) manifold.

\begin{figure*}[!htb]
    \centering
    \includegraphics[width=\textwidth]{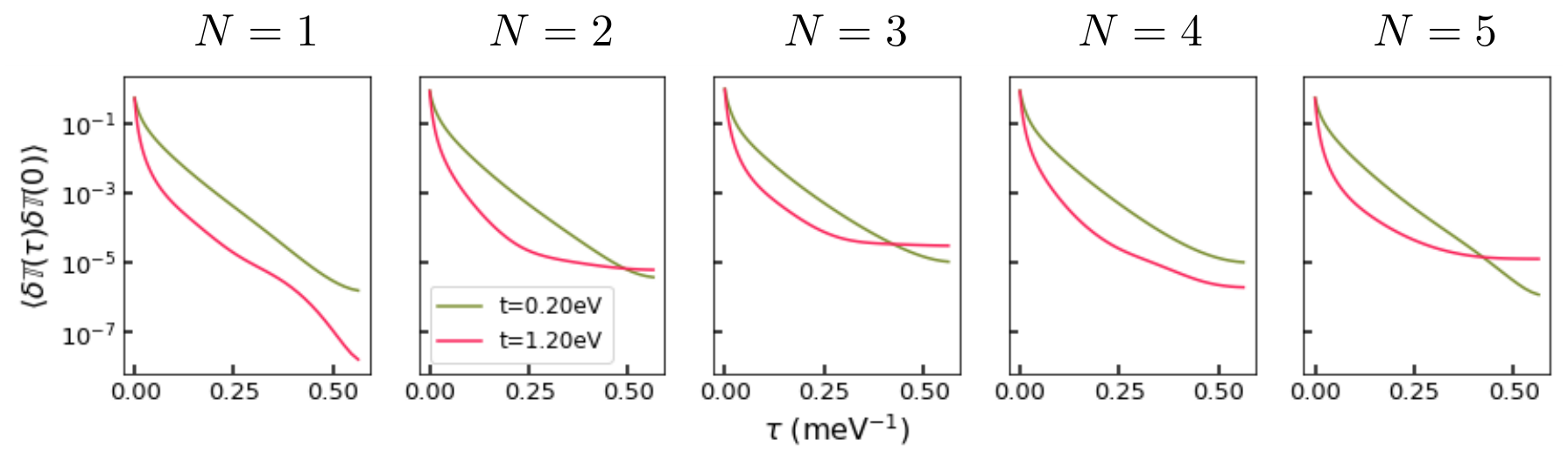}
    %\caption{    \textbf{Imaginary-time correlators for orbital and spin-orbital fluctuations.} 
    %Correlation functions $\langle \delta \mathbb{T}(\tau) \delta \mathbb{T}(0) \rangle$ (top) and $\langle \delta \mathbb{W}(\tau) \delta \mathbb{W}(0) \rangle$ (bottom) as a function of imaginary time $\tau$ for different hopping amplitudes ($t = 0.05$ eV, dashed black; $t = 0.2$ eV, green; $t = 1.2$ eV, red). The symmetric shape around $\tau = \beta/2$ reflects the bosonic nature of the correlators. Increasing hopping enhances the decay rate at short $\tau$, signaling the onset of faster orbital fluctuations and suggesting the emergence of gapped behavior at large $t$. Calculations performed at \( T = 10\,\mathrm{K} \) (\( \beta \approx 1160 \)), using parameters representative of 4d systems, \( g = B = 0.1\,\mathrm{eV} \), \( U = 2.5\,\mathrm{eV} \), and \( J = 0.4\,\mathrm{eV} \)}
    \caption{
Imaginary-time orbital–orbital and spin–orbital correlators for fillings \( N = 1 \) to \( N = 5 \) (left to right), computed for spin–orbit coupling \( \xi = 0.1 \)~eV, Jahn–Teller coupling \( g = 0.1 \)~eV, Hubbard interaction \( U = 2.5 \)~eV, and Hund's exchange \( J = 0.4 \)~eV. For each case, two hopping amplitudes are shown, with values \( t = 0.2 \)~eV (green) and \( t = 1.2 \)~eV (red). Due to the bosonic symmetry \( C(\tau) = C(\beta - \tau) \), the correlators are displayed for \( 0 \leq \tau \leq \beta/2 \). The more rapid decay and flattening at large \( \tau \) for \( t = 1.2 \)~eV indicate gapped excitations, while the smoother decay at \( t = 0.2 \)~eV reflects quasi-gapless dynamics. All calculations were performed at \( T = 10\,\mathrm{K} \), corresponding to an inverse temperature \( \beta = 1 / k_{\mathrm{B}} T \approx 1160\,\mathrm{eV}^{-1} \), using parameters representative of 4d systems: \( g = B = 0.1\,\mathrm{eV} \), \( U = 2.5\,\mathrm{eV} \), and \( J = 0.4\,\mathrm{eV} \).}
    \label{figtautau}
\end{figure*}

On the other hand, spin--spin correlations $\langle \delta S^z_{\vec{q}} \delta S^z_{\vec{-q}} \rangle$ exhibit a pronounced momentum structure that parallels the orbital--orbital and spin--orbital channels. As shown in Figure~\ref{fig3M}, these correlations tend to be suppressed at the $\Gamma$ point and enhanced at high-symmetry wavevectors such as $M = (\pi,\pi,0)$, $R = (\pi,\pi,\pi)$, and $X = (\pi,0,0)$. This pattern reflects a strong tendency toward staggered spin alignment and the emergence of antiferromagnetic-like textures. Importantly, this enhancement is reinforced by increasing the hopping amplitude. For values $t = 0.2$--$1.2$~eV, the sharpening of peaks indicates that intersite hybridization promotes the development of antiferromagnetic correlations, consistent with superexchange-driven antiferromagnetic tendencies known to arise in correlated $t_{2g}$ systems~\cite{khomskii2020orbital, takayama2021spin}. In contrast, at weak hopping ($t = 0.05$~eV, Figure~\ref{fig3M}b), the correlations become flat across momentum space, signaling a localized regime where fluctuation propagation is strongly suppressed.

\section{Imaginary-Time Dynamics}

\begin{figure*}%[ht] 
    \centering
    \includegraphics[width=0.6\textwidth]{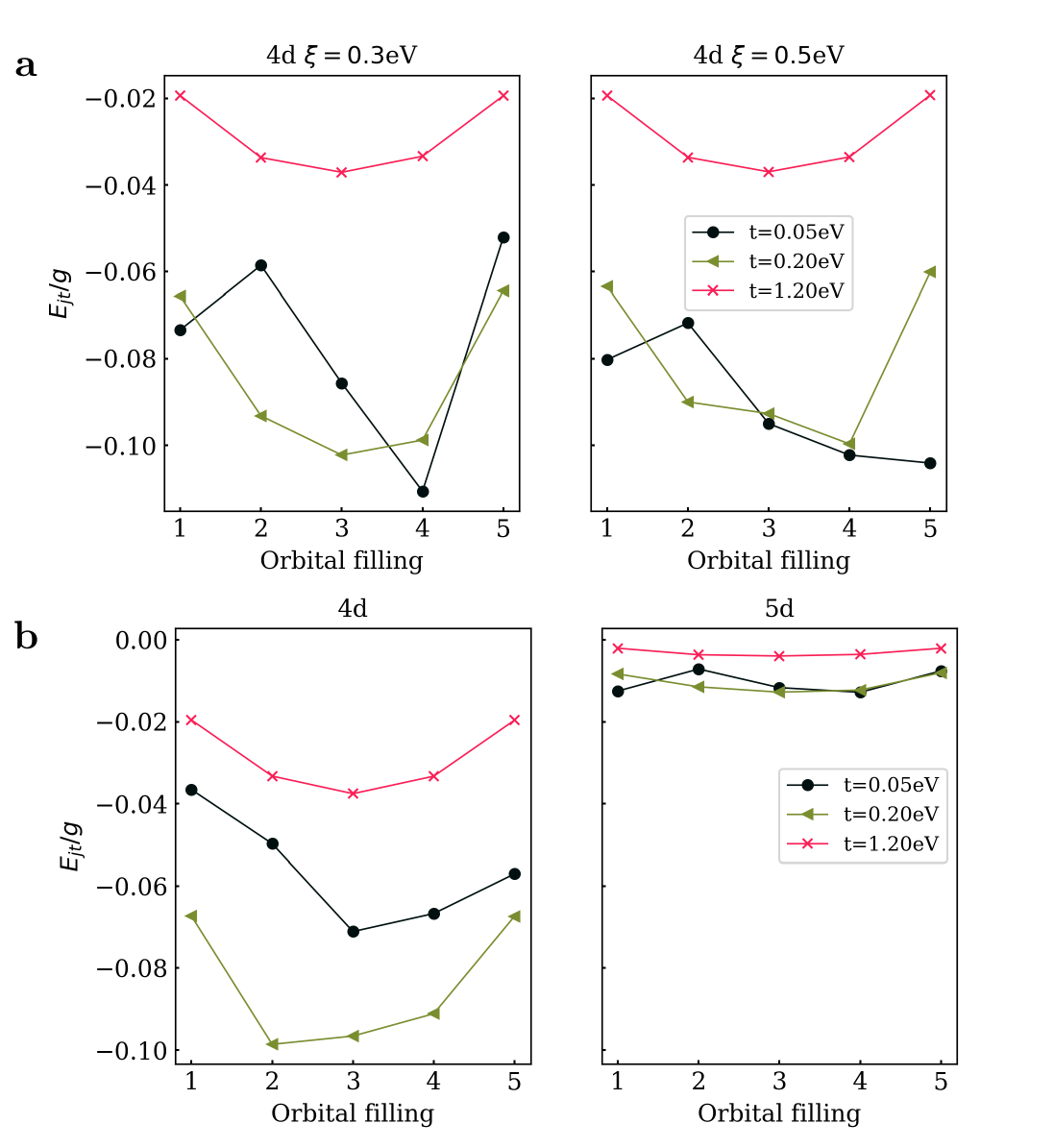} % Replace with your image file name
\caption{Energetic stabilization of orbital--lattice interactions.  
(a) Momentum-averaged JT energy $E_{jt}$ as a function of orbital filling \(N\) for two values of the spin-orbit coupling $\xi$ = 0.3 and 0.5 eV, demonstrating the relative insensitivity of orbital-lattice coupling to SOC. Calculations used parameters representative of 4$d$ systems: $U = 2.5$ eV, $J = 0.4$ eV, spin--orbit coupling $\xi = 0.1$ eV, and Jahn--Teller coupling $g = 0.1$ eV. (b) Momentum-averaged energy \(E_{\text{jt}}/g\) as a function of orbital filling \(N\) for two representative cases: 4$d$ (left, \(g = 0.1\,\mathrm{eV}\)) and 5$d$ (right, \(g = 0.01\,\mathrm{eV}\)). The reduced energy gain in 5$d$ systems reflects their weaker orbital-lattice coupling. Nonetheless, orbital polarization emerges as a result of local perturbations in both cases (see Figure \ref{Fig2M}), demonstrating that its emergence is primarily governed by electronic hopping and hybridization. Therefore, these results highlight that short-range orbital polarization can arise from intersite hopping alone, even in the absence of significant static Jahn-Teller coupling.}
\label{fig7new}
\end{figure*}

To probe the low-energy dynamics of orbital and spin-orbital correlations induced by the local perturbations, we analyze the imaginary-time correlators, which provide a useful framework for exploring the dynamical properties of many-body systems~\cite{mcardle2019variational, chang2024imaginary, shankar2017quantum}. Specifically, we examine \(\langle \delta \mathbb{T}(\tau) \delta \mathbb{T}(0) \rangle\) and \(\langle \delta \mathbb{W}(\tau) \delta \mathbb{W}(0) \rangle\) across fillings \(N = 1\)–\(5\) (Figure~\ref{figtautau}). In imaginary time, correlators for bosonic operators are defined as
\[
C(\tau) = -\langle \mathcal{T}_\tau A(\tau) A(0) \rangle,
\]
where \(\mathcal{T}_\tau\) denotes imaginary-time ordering and \(A(\tau) = e^{\tau \hat{H}} A e^{-\tau \hat{H}}\). These correlators are periodic and obey the symmetry
\[
C(\tau) = C(\beta - \tau),
\]
where \(\beta = 1 / k_{\mathrm{B}} T\) is the inverse temperature, with \(k_{\mathrm{B}}\) the Boltzmann constant and \(T\) the temperature. This relation follows from bosonic statistics and the Kubo–Martin–Schwinger condition~\cite{fetter1971theory}. Exploiting this symmetry, we restrict our plots in Figure~\ref{figtautau} to \(0 \leq \tau \leq \beta/2\) without loss of information. The long-\(\tau\) behavior of \(C(\tau)\) provides insight into the excitation spectrum, since a power-law decay is characteristic of gapless dynamics, while an exponential decay \[C(\tau) \sim e^{-\Delta \tau}\] signals the presence of a finite excitation gap \(\Delta\). Our data show that for \(t = 1.2\)~eV (red curves), the correlators (except for $N = 1$) decay rapidly at small \(\tau\) and flatten near \(\tau = \beta/2\) (Figure~\ref{figtautau}), suggesting the emergence of a finite gap in orbital and spin-orbital fluctuations. By contrast, for \(t = 0.2\)~eV (green curves), the decay is smoother and more gradual, consistent with gapless or weakly gapped dynamics. In the light of these results, it appears that large hopping amplitude \( t \)  suppresses low-energy fluctuations, whereas smaller \( t \) promotes local orbital dynamics and entangled spin-orbital modes, allowing for gapless excitations and slower decay in imaginary time. A definitive distinction between gapped and gapless regimes would require analytic continuation~\cite{abrikosov1963methods,mahan2000many,goulko2017numerical}, which would directly reveal the low-energy dynamics, enabling unambiguous identification of excitation gaps.

\section{Energetic stabilization and electron--lattice coupling}

To further confirm that the orbital polarization arises from electronic effects rather than from orbital-lattice interactions, we compute the Jahn--Teller stabilization energy \( E_{\text{jt}} \) using the Galitskii--Migdal formula~\cite{galitskii1958application,onida2002electronic}:
\begin{align}
    E_{\text{jt}} = -\operatorname{tr} \left[ \int_0^\beta d\tau \, \hat{G}(\tau) \, \hat{\Sigma}^{ep}(\beta - \tau) \right], \label{eint}
\end{align}
where \( \hat{G}(\tau) \) is the imaginary-time Green's function and \( \hat{\Sigma}^{ep}(\tau) \) is the electron--phonon self-energy. We evaluate \( E_{\text{jt}} \) as a momentum-averaged quantity over the Brillouin zone and study its dependence on the hopping amplitude \( t \), orbital filling \( N \), JT coupling strength \( g \), and spin--orbit coupling \( \xi \).

Figure~\ref{fig7new}a reveals that \( E_{\text{jt}} \) is only weakly sensitive to spin--orbit coupling. This relative insensitivity indicates that SOC plays little role in the energy gained from orbital-lattice interactions. On the other hand, Fig.~\ref{fig7new}b shows the normalized stabilization energy \( E_{\text{jt}}/g \) for representative 4$d$ and 5$d$ systems. As expected for 5$d$ systems, which have a significantly smaller JT coupling constant (\( g = 0.01\,\mathrm{eV} \)), the JT energy gain is much smaller than in 4$d$ systems (Fig.~\ref{fig7new}b). Yet, as discussed earlier, real-space orbital correlations persist even in this limit, confirming that short-range orbital polarization arises independently of lattice distortions. Together, these results demonstrate that intersite hopping alone suffices to restore orbital polarization in response to local perturbations. 

\section{Conclusions}

In summary, our analysis reveals that short-range orbital polarization emerges in response to local perturbations of the ground state in strongly spin-orbit-coupled systems. Notably, these effects arise even when the ground state is globally symmetric, with no net orbital polarization. The mechanism behind does not rely on orbital-lattice coupling but instead stems from an electronic effect by which orbital occupations are locally redistributed through intersite band hybridization in multiorbital systems. In perspective, analytic continuation of imaginary-time Green’s functions and self-energies to the real axis~\cite{abrikosov1963methods,mahan2000many,goulko2017numerical} may provide a viable route to access quasiparticle dynamics and transport properties. Integration with first-principles calculations could further extend this framework to material-specific predictions. In particular, advanced methods such as cluster DMFT~\cite{aoki2014nonequilibrium,schuler2020nessi,eckstein2021simulation,hariki2020lda+,murakami2023photo} are well suited to capture such short-range orbital correlations, as they explicitly incorporate spatial fluctuations beyond the single-site level.

In addition, structural distortions, such as octahedral tilts and rotations, can be incorporated via anisotropic hopping, allowing to add steric effects to the model. This may be relevant in systems where orbital polarization does not arise from conventional Jahn-Teller effects ~\cite{varignon2019origin}. Properly incorporating these contributions would allow for a more complete understanding of how lattice geometry mediates the competition between spin--orbit coupling and orbital-lattice interactions. Extensions to include trigonal or orthorhombic crystal fields~\cite{strestolvPhysRevX.10.031043,stretsolvtrigoPhysRevB.105.205142} and $t_{2g}$–$e_g$ orbital mixing~\cite{stamokostas2018mixing} may also reveal novel collective instabilities and hybrid spin--orbital excitations.

Interestingly, resonant inelastic X-ray scattering (RIXS) can conceptually be regarded as a type of quantum quench~\cite{ament2011resonant, nag2020observation, kim2022quantum, wang2023resonant}, bearing some resemblance to the local perturbations considered in our approach. While RIXS involves inherently nonperturbative processes, such as strong core-hole potentials and dipole transitions that lie beyond the scope of our model, its intrinsic orbital sensitivity makes it conceptually relevant to the type of fluctuations we study. Our results suggest that dynamic orbital polarization, driven by electronic hybridization, may contribute to the low-energy excitation spectrum, even if such contributions are difficult to resolve directly.

To estimate whether these fluctuations might fall within the range of current experimental sensitivity, we consider the characteristic energy scale associated with hybridization-driven orbital dynamics
\[
\Delta_{\mathrm{orbital}} \sim \frac{(Z t)^2}{U_{\mathrm{eff}} + \Delta_{\mathrm{SOC}}},
\]
where \( Z \) is the quasiparticle weight, \( t \) the hopping amplitude, \( U_{\mathrm{eff}} \) the effective Coulomb interaction, and \( \Delta_{\mathrm{SOC}} \) the spin-orbit splitting~\cite{khaliullin2005orbital, streltsov2017suppression, georges2013strong}. This estimate assumes that each hopping process is renormalized by a quasiparticle factor \( Z \). For typical 4$d$ and 5$d$ transition-metal oxides, we use representative values, i.e., \( t \sim 0.2\text{--}1\,\mathrm{eV} \)~\cite{streltsov2017suppression, hafez2024antiferromagnetic}, \( Z \sim 0.2\text{--}0.6 \)~\cite{georges2013strong}, and \( U_{\mathrm{eff}} + \Delta_{\mathrm{SOC}} \sim 2\text{--}3\,\mathrm{eV} \). These yield characteristic energy scales that range from a few meV to a few tens of meV, and therefore may be below the resolution limits of current high-resolution RIXS~\cite{ament2011resonant, moretti2011resonant}. Although such fluctuations are therefore unlikely to be directly observed in present-day RIXS spectra, they may still influence the low-energy lineshape through coupling with other excitations, or contribute to collective responses observable via indirect signatures. %These considerations are particularly relevant in 4$d$ systems, where somewhat larger hopping amplitudes and quasiparticle weights could shift the energy scale upward, albeit still near the limits of detectability.

Additionally, though still challenging, polarization-resolved RIXS measurements may provide a pathway to distinguish hybridization-driven orbital fluctuations from other low-energy excitations. Cross-polarized channels (such as \( \pi\text{--}\sigma \) or \( \sigma\text{--}\pi \)) are known to selectively probe transitions between orthogonal orbitals, and could, in principle, enhance sensitivity to the types of orthogonal orbital fluctuations described here.
Likewise, momentum-resolved RIXS experiments conducted at high-symmetry points (e.g., \( X \), \( M \), and \( R \)) may reveal weakly dispersive spectral features with characteristic polarization-dependent intensity modulations. While these techniques offer indirect routes to assess the presence of orbital dynamics, we emphasize that any quantitative predictions must be based on material-specific models incorporating realistic Hamiltonians, explicitly computed dipole matrix elements, core-hole effects, and orbital selection rules.\\

\noindent \textbf{Acknowledgments}\\ We wish to thank Kyle Shen, Ankit Disa and Craig J. Fennie for stimulating and valuable discussions. This work was supported by Projects No. PID2023-152225NB-100 and Severo Ochoa MATRANS42 (No. CEX2023-001263-S) of the Spanish Ministry of Science and Innovation (Grant No. MICIU/AEI/10.13039/501100011033 and FEDER, EU), Projects No. TED2021-129857B-I00 and PDC2023-145824-I00, funded by MCIN/AEI/10.13039/501100011033 and European Union NextGeneration EU/PRTR, and the Generalitat de Catalunya (2021 SGR 00445).

\appendix
\section{Self-consistent equations} \label{app_A}
\renewcommand{\theequation}{A\arabic{equation}}

\noindent
To solve the lattice Hamiltonian, we employed the following self-consistent equations:

\vspace{1em}

%\begin{widetext}
\begin{align}
\begin{split}
\hat{G}_\vec{k}(\imath\omega_n) &= 
\left\{
\hat{\gfree}^{-1}_\vec{k}(\imath\omega_n) 
- \left[
\hat{\Sigma}^{HF} 
+ \hat{\Sigma}^{(2)}(\imath\omega_n) \right. \right. \\ &\phantom{=}
\left. \left. \hspace{60pt}
+ \hat{\Sigma}^{ep}(\imath\omega_n)
+ \hat{\Sigma}^{oex}_\vec{k}
\right] \right\}^{-1} \label{eq:G}
\end{split}\\
D_\eta(\imath\nu_n) &= 
\left[
\freephon^{-1}_\eta(\imath\nu_n) 
- \Pi_\eta(\imath\nu_n) \right]^{-1} \label{eq:D}
\end{align}
%\end{widetext}
%
where, for a given reciprocal vector $\vec{k}$, $\hat{G}_\vec{k}(\imath\omega_n)$ is the dressed electron Green's function and $\hat{\gfree}^{-1}_\vec{k}(\imath\omega_n)=[i\omega_n+\mu-\varepsilon_0-\varepsilon_k]$ is the bare Green’s function, with $\mu$ the chemical potential and $\varepsilon_0, \varepsilon_k$ are the eigenvalues of the on-site and kinetic terms of the Hamiltonian lattice without interactions. The terms $\hat{\Sigma}^{HF}$ and $\hat{\Sigma}^{(2)}(\imath\omega_n)$ correspond to the Hartree-Fock and second-order Born approximations to the electronic self-energy, while $\hat\Sigma^{oex}_\vec{k}$ and $\hat{\Sigma}^{ep}(\imath\omega_n)$ denote the orbital exchange and electron-phonon self-energy elements, respectively. The self-energy expressions in terms of Green’s functions are provided in Appendix \ref{MatInv}. Fermionic and bosonic frequencies are represented by $\imath\omega_n$ and $\imath\nu_n$, respectively. The self-energies without the subscript \( \vec{k} \) are independent of reciprocal vectors due to their local nature, while $\hat{\Sigma}^{HF}$ and $\hat\Sigma^{oex}_\vec{k}$ are local in imaginary time and, in consequence, do not depend on the fermionic frequency $\imath\omega_n$. Additionally, \(\freephon^{-1}_\eta(\imath\nu_n)\) and \(D_\eta(\imath\nu_n)\) correspond to the free and dressed phonon propagators, while \(\Pi_\eta(\imath\nu_n)\) denotes the phonon self-energy (Equation (4) of the main text).\\

\begin{figure*}[htb]
    \centering
    \includegraphics[width=\textwidth]{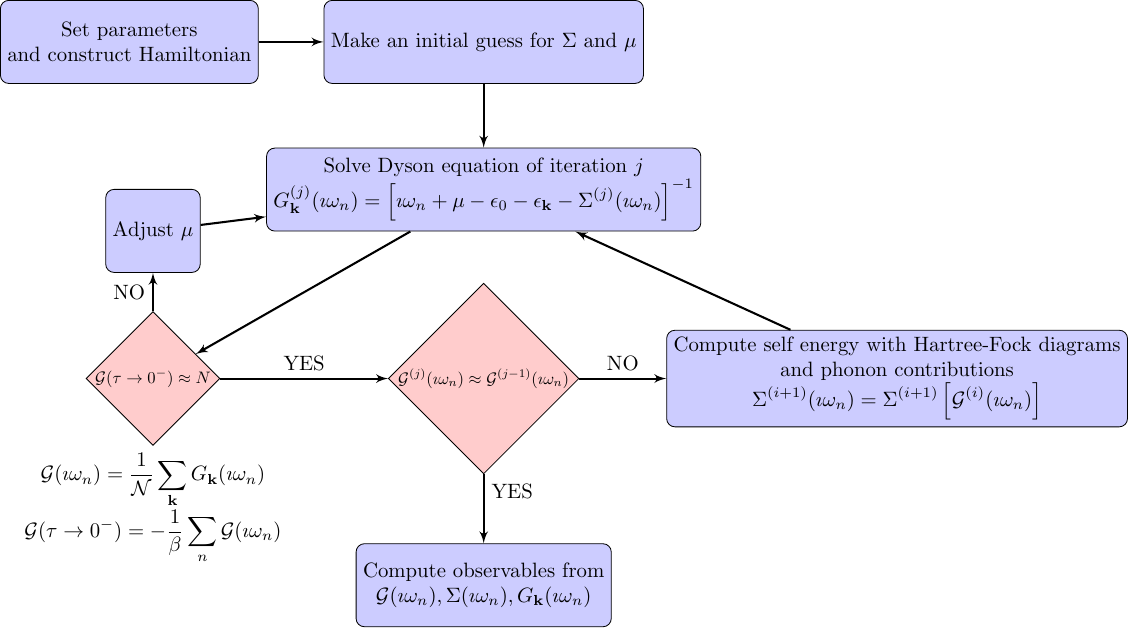} % Adjust the file name as needed
    \caption{
Workflow of the Dyson solver used in this study. The calculation begins by setting model parameters (temperature, interaction strengths, hopping amplitude, etc.) and constructing the lattice Hamiltonian. An initial guess is made for the self-energy (\( \Sigma(i\omega) \)) and chemical potential (\( \mu \)). The Dyson equation is then solved to compute the interacting Green's function (\( G(i\omega) \)), with iterative updates of \( \Sigma(i\omega) \) based on Hartree-Fock diagrams and phonon contributions. The particle density is checked for consistency with the input, and \( \mu \) is adjusted if necessary. Convergence is evaluated by comparing \( G(i\omega) \) across iterations. Once convergence is achieved, the results, including \( G(i\omega) \), \( \Sigma(i\omega) \), and derived observables (e.g., total energy and correlation functions), are saved, marking the end of the calculation.
    }
    \label{fig:dyson_workflow}
\end{figure*}

To accelerate the convergence of these equations, we employed the Direct Inversion in the Iterative Subspace (DIIS) method, which is described in Appendix \ref{DIIS}. Once the self-consistent equations are solved, we determined the average filling $N$ in the \( t_{2g} \) orbitals from the converged Matsubara Green functions (see Appendix~\ref{IRbasis}). This is achieved by taking the trace of the Matsubara Green function:
\begin{align}
   N = \langle n \rangle = -\operatorname{Tr} \left[ \hat{\gloc}(\beta) \right],
\end{align}
where the trace is taken over the orbital and spin degrees of freedom from the local Green's function
\begin{align}
   N = \langle n \rangle = -\operatorname{Tr} \left[ \hat{\gloc}(\beta) \right], \label{eq:particle_number}
\end{align}

\begin{figure*}[htb] % Use 'figure*' for a two-column wide figure
    \centering
    \includegraphics[width=0.8\textwidth]{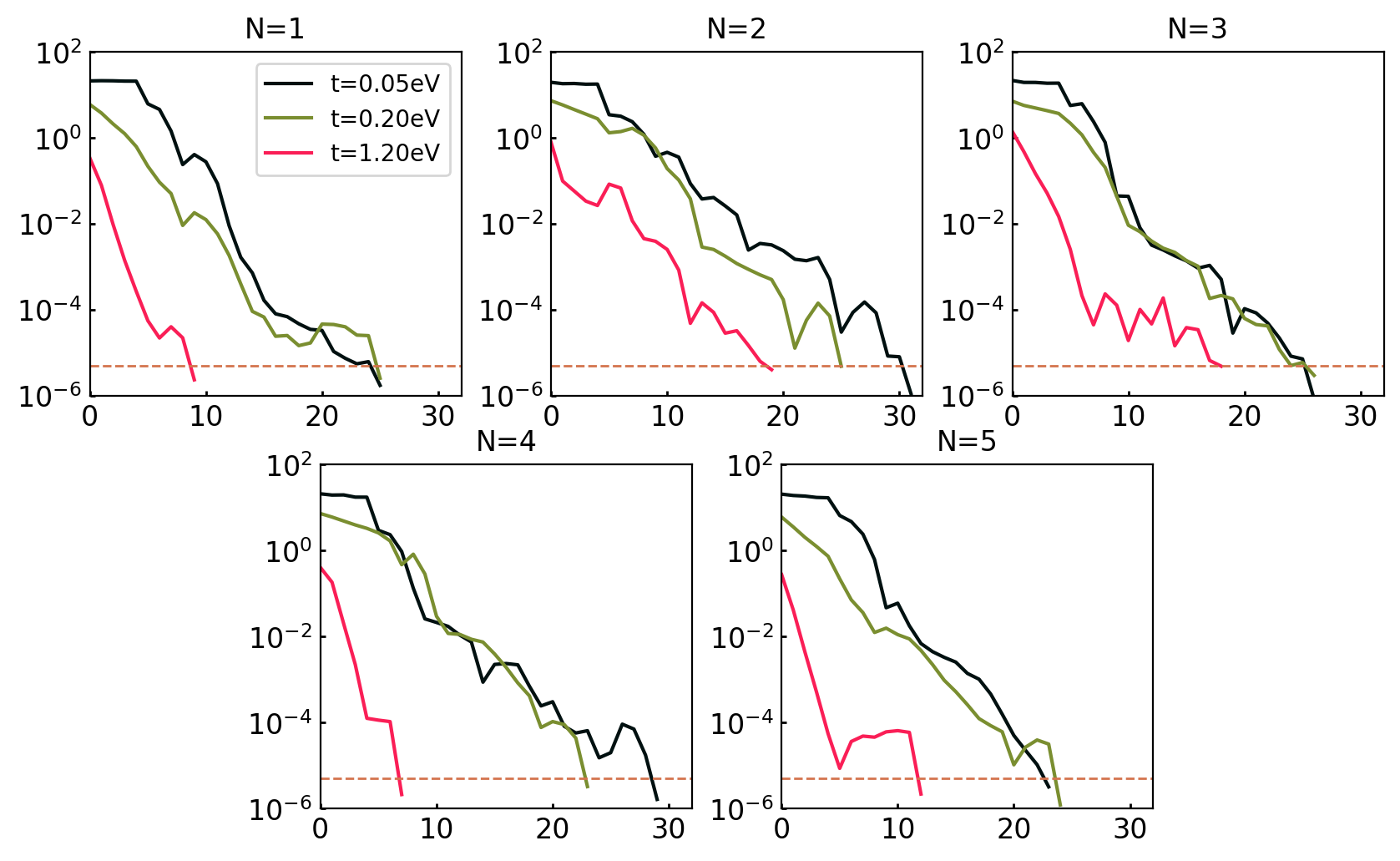} % Adjust width as needed for the two-column layout
\caption{
Convergence of the Matsubara Green functions \(G(i\omega)\) for different orbital fillings \(N \) and hopping amplitudes $t$. The horizontal axis represents the number of iterations. Calculations were performed at a temperature \(T = 10 \, \text{K}\), spin-orbit coupling strength \(\xi = 0.1 \, \text{eV}\), Coulomb interaction \(U = 2.5 \, \text{eV}\), and Hund's coupling \(J = 0.4 \, \text{eV}\). The dashed orange line marks the numerical threshold (\(5\cdot 10^{-6}\)) for convergence using the DIIS algorithm.
}
    \label{fig4conv} % Replace 'example_label' with a meaningful label
\end{figure*}

The local Green's function, written in imaginary frequency, is $\hat{\gloc}(\imath \omega_n) = \frac{1}{\mathcal{N}}\sum_k G_k(\imath \omega_n)$, where $\mathcal{N}$ is the number of grid points in reciprocal space. The chemical potential \( \mu \) is adjusted iteratively to ensure that the orbital filling \( N \) reaches the target value within the specified convergence threshold. Figure \ref{fig:dyson_workflow} shows the workflow chart used to solve self-consistent equations. Figure \ref{fig4conv} shows representative examples of convergence of Matsubara Green functions \(\hat{G}(\imath\omega_n)\) for different orbital occupancies and hopping amplitudes.

\section{Direct inversion in the iterative subspace (DIIS) algorithm} \label{DIIS}
\renewcommand{\theequation}{B\arabic{equation}}
% References of this section comes from https://doi.org/10.1063/5.0082586

Direct inversion in the iterative subspace (DIIS) is an extrapolation technique designed to accelerate the convergence of self-consistent methods \cite{pokhilko2022iterative, kudin2002black}. The goal is to determine a vector $\vec{v}^*$ through successive self-consistent iterations, labeled by $\vec{v}^{(k)}$. At each iteration, the error vector $\vec{e}^{(k)} = \vec{v}^* - \vec{v}^{(k)}$ is computed. DIIS minimizes this error by forming a linear combination of vectors from the iterative subspace
\begin{align}
    \vec{e}^{ext} &= \vec{v}^* - \sum_k c_k \vec{v}^{(k)}.
\end{align}
Since the solution vector is unique, the coefficients must satisfy the constraint $\sum_k c_k = 1$. The minimization of $|\vec{e}^{ext}|$ with this constraint is expressed through the Lagrange multiplier method
\begin{align}
    \mathcal{L}(\vec{c}, \lambda) &= \frac{1}{2}\sum_{i,j} c_i c_j B_{ij} - \lambda \left( 1 - \sum_i c_i \right),
\end{align}
where $B_{ij} = \langle \vec{e}^{(i)}, \vec{e}^{(j)} \rangle$, denoting an inner product (more details below). Minimizing $\mathcal{L}$ reduces to solving the linear algebra problem
\begin{align}
    \begin{pmatrix}
        B & \vec{1} \\ \vec{1}^T & 0
    \end{pmatrix} 
    \begin{pmatrix}
        \vec{c} \\ \lambda
    \end{pmatrix} = 
    \begin{pmatrix}
        \vec{0} \\ 1
    \end{pmatrix}.
\end{align}
Two practical challenges arise. First, $\vec{v}^*$ is not known \textit{a priori}, requiring approximate error computation. A common approach is to use $\vec{e}^{(k)} = \vec{v}^{(k)} - \vec{v}^{(k-1)}$, which approaches zero as convergence improves. Second, the iterative subspace dimension grows with iterations. This is managed by truncating DIIS to a fixed number of past iterations.

In our case, we minimize the iterative error of the self-energy vector:
\begin{align}
    \vec{v} = \left[\hat\Sigma^{HF}, \hat\Sigma^{(2)}(\tau)\right],
\end{align}
where the dependence on $\tau$ (except for $\hat\Sigma^{HF}$, which is time-local) refers to a finite imaginary time grid determined by the intermediate representation (IR) basis, which is described in Appendix \ref{IRbasis}. Each error vector component is a matrix, and the inner product used in the minimization step is defined using the Frobenius norm
\begin{align}
    \langle \vec{e}^{(i)}, \vec{e}^{(j)} \rangle &= \sum_r \text{tr}\left[ \hat{e}^{(i)}_r \hat{e}^{(j)}_r \right].
\end{align}
The Frobenius norm is specifically employed in forming the matrix $B_{ij}$ during the minimization step, where the iterative coefficients $c_k$ are determined. This ensures a consistent and meaningful measure of error magnitude between different matrix-valued components of the self-energy, making it critical in the linear algebraic step of DIIS that solves for the optimal coefficients.

\section{Intermediate representation (IR) basis} \label{IRbasis}
\renewcommand{\theequation}{C\arabic{equation}}

Transforming between imaginary time (\(\tau\)) and imaginary frequency (\(i\omega_n\)) is advantageous because Dyson equations are most straightforwardly solved in frequency space, where convolutions simplify to multiplications. Since many self-energy components (\(\hat{\Sigma}^{(2)}(\tau)\), \(\hat{\Sigma}^{ep}(\tau)\)) and Green's functions are naturally expressed in imaginary time, accurate and efficient Fourier transformations between these two representations become essential in solving the self-consistent Dyson equations. The intermediate representation (IR) basis ensures this transformation can be done efficiently~\cite{shinaoka2017compressing, li2020sparse, wallerberger2023sparse}. The IR approach introduces a kernel \(K(\tau, \omega)\) that directly relates imaginary time and imaginary frequency domains
\begin{equation}
G_{\mu\nu}(\tau) = -\int_{-\omega_M}^{\omega_M} d\omega \, \frac{\mathrm{e}^{-\omega \tau}}{1 \mp \mathrm{e}^{-\omega \beta}} A_{\mu\nu}(\omega).
\label{eq:green_tau}
\end{equation}
where $G_{\mu\nu}(\tau)$ represents a matrix element of the electron dressed Green's function for orbitals $\mu, \nu$  and 
$A_{\mu\nu}(\omega)$ the corresponding spectral functions. The kernel \(K(\tau, \omega) \) can be efficiently decomposed using singular value decomposition (SVD) as
\begin{equation}
K(\tau, \omega) = \sum_{l=0}^{\infty} U_l(\tau) S_l V_l(\omega),
\label{eq:kernel_svd}
\end{equation}
where the rapidly decaying singular values \(S_l\) allow for a highly accurate truncation. In this representation, the Green's function and self-energy components are expanded in the IR basis as
\begin{align}
\hat{G}_l &= \int_0^\beta U_l(\tau) \hat{G}(\tau) \, d\tau,
\label{eq:green_basis} \\
\hat{\Sigma}_l &= \int_0^\beta U_l(\tau) \hat\Sigma(\tau) \, d\tau.
\label{eq:sigma_basis}
\end{align}

To perform the Fourier transform from imaginary time to Matsubara frequency, one first projects the imaginary time function onto the IR basis as in Eq. \eqref{eq:green_basis} or Eq. \eqref{eq:sigma_basis}. Then, using the precomputed Fourier transforms of the IR basis functions \(\hat{U}_l(i\omega_n)\) by the kernel definition into imaginary frequency, $K(\imath\omega_n,\omega) = (\imath\omega_n-\omega)^{-1}$. Then, we have:
\begin{align}
\hat{G}(i\omega_n) = \sum_l U_l(i\omega_n) \hat{G}_l,
\label{eq:green_tau_to_omega}\\
\hat\Sigma(i\omega_n) = \sum_l U_l(i\omega_n) \hat\Sigma_l.
\label{eq:sigma_tau_to_omega}
\end{align}
Conversely, to transform back from imaginary frequency to imaginary time, one uses the inverse transformation:
\begin{align}
\hat{G}(\tau) &= \sum_l U_l(\tau) \hat{G}_l, && \text{where} & \hat{G}_l &= \sum_n U_l(i\omega_n) \hat{G}(i\omega_n),
\label{eq:green_omega_to_tau}\\
\hat\Sigma(\tau) &= \sum_l U_l(\tau) \hat\Sigma_l, && \text{where} & \hat\Sigma_l &= \sum_n U_l(i\omega_n) \hat\Sigma(i\omega_n).
\label{eq:sigma_omega_to_tau}
\end{align}

In order to perform these transformations it is necessary to make a grid on imaginary time and a truncation on imaginary frequency~\cite{li2020sparse}. This is precomputed in the Python module \texttt{sparse\_ir} where piecewise Legendre polynomials are used to define $U_l(\tau)$~\cite{shinaoka2017compressing}. When \( N \) singular values are needed to accurately capture the precision of the Green's function (for \( l = 0, \ldots, N - 1 \)), the zeros of the \( N \)th polynomial are employed for \( \tau \)-sampling. Then, $U_l(\imath\omega_n)$ are related to spherical Bessel functions, which are the Fourier transforms of Legendre polynomials. Consequently, the sampling and truncation in the imaginary frequency domain are determined by the zeros of the $N$-th spherical Bessel function. Since Matsubara frequencies are discrete and do not always coincide with these zeros, the nearest available value is chosen.

\section{Propagators and self-energies} \label{MatInv}
\renewcommand{\theequation}{D\arabic{equation}}

In the following, we give explicit expressions for the self-energies and propagators discussed in Equation \ref{eq:G}. To improve computational efficiency, it is advantageous to separate the electronic propagators and self-energies into diagonal and off-diagonal components. We therefore start with the general discussion of the algebraic properties of matrices that exhibit this structure, and particularly why are they useful for computational efficiency and subsequently we demonstrate how Green's functions and self-energy matrices do indeed display such matrix structure.

\subsection{Increasing computational efficiency through matrix decomposition into diagonal and off-diagonal components}

To solve the self-consistent equations, we analyze matrices of the form
\begin{equation}
\mathcal{\hat M} = a\hat{\mathbb{I}} + b\mathcal{\hat V},
\label{eq:matrix_form}
\end{equation}
where \(a\) and \(b\) are scalars, \(\hat{\mathbb{I}}\) is the identity matrix, and \(\mathcal{\hat V}\) is a matrix with the property
\begin{equation}
\mathcal{\hat V}^2 = 2\hat{\mathbb{I}} + \mathcal{\hat V}.
\label{eq:matrix_property}
\end{equation}
This algebraic structure of \(\mathcal{\hat V}\) allows us to derive compact expressions for key operations, such as finding the inverse and the exponential of \(\mathcal{\hat M}\). To compute the inverse of \(\mathcal{\hat M}\), we assume it has a similar form
\begin{equation}
\mathcal{\hat M}^{-1} = c\hat{\mathbb{I}} + d\mathcal{\hat V},
\label{eq:inverse_form}
\end{equation}
where \(c\) and \(d\) are unknown coefficients to be determined. Using the property of matrix inversion, \(\hat{\mathbb{I}} = \mathcal{\hat M}\mathcal{\hat M}^{-1}\), and substituting the assumed forms of \(\mathcal{\hat M}\) and \(\mathcal{\hat M}^{-1}\), we expand the product
\begin{equation}
\hat{\mathbb{I}} = (ac + 2bd)\hat{\mathbb{I}} + (ad + bc + bd)\mathcal{\hat V}.
\label{eq:inverse_product}
\end{equation}
By equating the coefficients of \(\hat{\mathbb{I}}\) and \(\mathcal{\hat V}\), we obtain two equations
\begin{equation}
ac + 2bd = 1, \quad ad + bc + bd = 0.
\label{eq:coefficients_equations}
\end{equation}
Solving this system of equations, we find:
\begin{equation}
c = \frac{a + b}{a^2 - 2b^2 + ab}, \quad d = -\frac{b}{a^2 - 2b^2 + ab}.
\label{eq:coefficients_solutions}
\end{equation}
Thus, the inverse of \(\mathcal{\hat M}\) is given by:
\begin{equation}
\mathcal{\hat M}^{-1} = \frac{(a + b)\hat{\mathbb{I}} - b\mathcal{\hat V}}{a^2 - 2b^2 + ab}.
\label{eq:inverse_result}
\end{equation}

This result shows that the matrix $\mathcal{\hat M}$ and its inverse $\mathcal{\hat M}^{-1}$ retain the same structural properties, enabling efficient matrix operations. More importantly, this framework extends naturally to the computation of any function $f(\mathcal{\hat M})$, such as $\euler{\alpha\mathcal{\hat M}}$ (with $\alpha \in \mathbb{C}$) or $\sin(\omega\mathcal{\hat M})$. In particular, the eigenvalues of \(\mathcal{\hat M}\) are (\(a - b\)) with multiplicity 4 and (\(a + 2b\)) with multiplicity 2. These eigenvalues allow us to express any arbitrary function of \(\mathcal{\hat M}\) in the form  
\begin{equation}
f({\mathcal{\hat M}}) = c\hat{\mathbb{I}} + d\mathcal{\hat V},
\label{eq:func_form}
\end{equation}
where \(c\) and \(d\) are coefficients to be determined. Using the eigenvalue relationships, one can impose the conditions  
\begin{align}
    f(a-b) = c-d, \quad f(a+2b) = c+2d.
    \label{eq:func_coefficients}
\end{align}
Solving this system of equations, we obtain  
\begin{equation}
c = \frac{f(2a+b) + 2f(a-b)}{3}, \quad d = \frac{f(2a+b) - f(a-b)}{3}.
\label{eq:func_solutions}
\end{equation}

These results relate the structure of \(\mathcal{\hat M}\) with its inverse and every arbitrary function. The derived expressions highlight how the properties of \(\mathcal{\hat V}\) simplify complex operations, providing a practical way for working with matrices of this form. In the following sections we explain how these properties can be exploited for the efficient computation of propagators and self-energies.

\subsection{Kinetic terms and spin-orbit coupling}

As aforementioned, separating the electronic propagators and self-energies into diagonal and off-diagonal components is advantageous to improve computational efficiency. We start the discussion of this point by neglecting electron-electron and electron-phonon interaction and focusing on the kinetic terms and spin-orbit coupling. In this scenario, the Hamiltonian can be expressed as
\begin{align} \label{eqn_hamdiag}
    \mathcal{\hat H} &= \varepsilon \hat{\mathbb{I}} + \dfrac{\xi}{2}\hat{\mathcal{V}},
\end{align}
The term \( \varepsilon \hat{\mathbb{I}} \) represents the kinetic energy in the cubic lattice, which, for simplicity, we assume is diagonal in the \( t_{2g} \) basis in \( \vec{k} \)-space, hence the identity matrix $\hat{\mathbb{I}}$. Off-diagonal contributions arise from spin-orbit coupling, described by the matrix \( \hat{\mathcal{V}} \) with coupling constant \( \frac{\xi}{2} \). The Matsubara non-interacting propagator is then given in matrix form by:
\begin{align} \label{eqn_diag}
\hat{\gfree}_{\vec{k}}(\imath\omega_n) &= \left[ (\imath\omega_n - \varepsilon_{\vec{k}} + \mu)\hat{\mathbb{I}} - \frac{\xi}{2}\hat{\mathcal{V}} \right]^{-1} \nonumber \\
&= \mathfrak{g}_{d,\vec{k}}(\imath\omega_n)\hat{\mathbb{I}} - \mathfrak{g}_{od,\vec{k}}(\imath\omega_n)\hat{\mathcal{V}}.
\end{align}
Here, \(\varepsilon_{\vec{k}}\) denotes the kinetic energy in the space \(\vec{k}\), \(\mu\) is the chemical potential, and \(\mathfrak{g}_{d,\vec{k}}(\imath\omega_n)\hat{\mathbb{I}}\) and \(\mathfrak{g}_{od,\vec{k}}(\imath\omega_n)\hat{\mathcal{V}}\) represent the diagonal and non-diagonal matrix components, respectively, where $\mathfrak{g}_{d,\vec{k}}(\imath\omega_n)$ and $\mathfrak{g}_{od,\vec{k}}(\imath\omega_n)$ are complex numbers that depend on $\imath\omega_n$ and momentum $\vec{k}$. As discussed above, this feature simplifies the inversion of matrices, which is essential for efficient computation of propagators and self-energies. Let us assume for the moment that both electron and phonon self-energy matrices can be similarly decomposed
\begin{equation} \label{eqn_diagforsigma}
    \hat\Sigma(\imath\omega_n) = \mathfrak{s}_d(\imath\omega_n) \hat{\mathbb{I}} - \mathfrak{s}_{od}(\imath\omega_n) \hat{\mathcal{V}},
\end{equation}
where $\hat\Sigma(\imath\omega_n)$ is a generic self-energy matrix, and $\mathfrak{s}_d(\imath\omega_n)$, $\mathfrak{s}_{od}(\imath\omega_n)$ denote complex numbers whose values depend on the imaginary frequency. Under this condition, the dressed Green's function matrix \(\hat{G}_\vec{k}(\imath\omega_n)\) retains the same structure as the bare propagator matrix \(\hat\gfree_{\vec{k}}\) in Eq.~\ref{eqn_diag}, giving
\begin{align} \label{eqn_fullG}
\hat{G}_\vec{k}(\imath\omega_n) &= \left[ (\imath\omega_n - \varepsilon_\vec{k} - \mathfrak{s}_d(\imath\omega_n) + \mu)\hat{\mathbb{I}} \right. \nonumber \\
&\quad \left. - \left(\frac{\xi}{2} + \mathfrak{s}_{od}(\imath\omega_n)\right)\hat{\mathcal{V}} \right]^{-1} \nonumber \\
&= \mathrm{g}_{d,\vec{k}}(\imath\omega_n)\hat{\mathbb{I}} - \mathrm{g}_{od,\vec{k}}(\imath\omega_n)\hat{\mathcal{V}}.
\end{align}

In the following, we validate this assumption, demonstrating that electron and electron-phonon self-energies retain the structure outlined in Eq.~\ref{eqn_diagforsigma}.

\subsection{Propagators and self-energies for Jahn-Teller phonons and vibronic interactions}

It is convenient to rewrite the Jahn-Teller coupling using Gell-Mann matrices, $\hat{\lambda}^\eta$, as follows:
\begin{align}
    \hamil_{JT} &= \sum_{i,\eta,\alpha,\beta,\sigma,\sigma'} g_\eta \lambda^\eta_{\alpha\beta} c_{i\alpha\sigma}^\dagger c_{i\beta\sigma'} \left( b_{i\eta} + b_{i\eta}^\dagger \right)
\end{align}
where $\hat{\lambda}^3$ and $\hat{\lambda}^8$ are Gell-Mann matrices related with $E_g$ Jahn-Teller distortion modes
\begin{subequations}
\begin{align} 
\hat{\lambda}^3 &= \begin{pmatrix}
1 & 0 & 0 \\ 0 & -1 & 0 \\
0 & 0 & 0 
\end{pmatrix} \\ \
\hat{\lambda}^8 &= \dfrac{1}{\sqrt{3}}\begin{pmatrix}
1 & 0 & 0 \\ 0 & 1 & 0 \\
0 & 0 & -2  
\end{pmatrix}
\end{align}
\end{subequations}
Using this formulation, the coupling constants of the two modes are \( g_\eta = g/\sqrt{3} \) for \( \eta = 3, 8 \)—where \( \eta = 3 \) corresponds to the orthorhombic \( \phi \) (\( Q_2 \)) mode, and \( \eta = 8 \) corresponds to the tetragonal \( \theta \) (\( Q_3 \)) mode. For all other \( \eta \) indices, the coupling constants are set to zero, i.e., \( g_\eta = 0 \). Then, using the charge-orbital moment \( T_{i\eta} = \sum_{\gamma,\gamma',\sigma} \lambda^\eta_{\gamma\gamma'} c_{i\gamma\sigma}^\dagger c_{i\gamma'\sigma} \) \cite{kaga2022eliashberg}, the dressed phonon propagator can be written in imaginary time \( \tau \) as follows:
\begin{widetext}
\begin{equation} \label{eqn_PI}
\begin{aligned}
\Pi_\eta(\tau) &= -|g_\eta|^2 \langle \timeorder T_{i\eta}(\tau) T_{i\eta}(0) \rangle \\
&= -|g_\eta|^2 \sum_{\substack{\gamma_1, \gamma'_1, \sigma_1 \\ \gamma_2, \gamma'_2, \sigma_2}} 
\lambda^\eta_{\gamma_1 \gamma'_1} \lambda^\eta_{\gamma_2 \gamma'_2} \,
\langle \timeorder\, c_{i \gamma_1 \sigma_1}^\dagger(\tau) c_{i \gamma'_1 \sigma_1}(\tau)
c_{i \gamma_2 \sigma_2}^\dagger c_{i \gamma'_2 \sigma_2} \rangle \\
&= -|g_\eta|^2 \sum_{\substack{\gamma_1, \gamma'_1, \sigma_1 \\ \gamma_2, \gamma'_2, \sigma_2}} 
\lambda^\eta_{\gamma_1 \gamma'_1} \lambda^\eta_{\gamma_2 \gamma'_2} \left\{
\gloc_{\gamma'_1 \sigma_1, \gamma_1 \sigma_1}(\tau) \gloc_{\gamma'_2 \sigma_2, \gamma_2 \sigma_2}(\tau)
- \gloc_{\gamma'_1 \sigma_1, \gamma_2 \sigma_2}(\tau)
\gloc_{\gamma'_2 \sigma_2, \gamma_1 \sigma_1}(-\tau) \right\} \\
&= \frac{4}{3} g \left[ \mathsf{g}_d(\tau) \mathsf{g}_d(-\tau)
- \mathsf{g}_{od}(\tau) \mathsf{g}_{od}(-\tau) \right].
\end{aligned}
\end{equation}
\end{widetext}
Here, $\mathrm{g}_d(\tau)$ and $\mathrm{g}_{od}(\tau)$ are real functions representing the components of the local Green's function $\hat{\mathcal{G}}(\tau)$, expressed in imaginary time instead of imaginary frequency. The operator $\mathcal{T}$ denotes time-ordering, and $\langle \dots \rangle$ represents the expectation value. Note, in particular, that in the final expression the self-energy of the phonon does not depend on $\eta$, so it can be immediately seen that $D_\theta(\imath\nu_n) = D_\phi(\imath\nu_n)$. Now we can express the dressed phonon propagator as
\begin{equation}
    D_\eta(\imath\nu_n) = \frac{1}{{\freephon}^{-1}_\eta(\imath\nu_n) - \Pi_\eta(\imath\nu_n)}
\end{equation}
where $\imath\nu_n$ are bosonic Matsubara frequencies and $\mathfrak{D}_\eta(\imath\nu_n) = 2\omega_{\eta,0}[(\imath\nu_n)^2-\omega_{\eta,0}^2]^{-1}$ is the bare bosonic propagator. Using the Migdal approximation \cite{migdal1958interaction, mahan2000many, giustino2017electron} and the dressed phonon propagator \(D_\eta(\imath\nu_n)\), we can derive the electron-phonon self-energy, which, expressed in imaginary time is
\begin{align} 
    \Sigma_{\mu\sigma,\nu\sigma'}^{ep}(\tau) &= \sum_{\eta,\gamma,\gamma'} |g_\eta|^2 D_\eta(\tau) \lambda^\eta_{\mu\gamma} \gloc_{\gamma\sigma,\gamma'\sigma'}(\tau) \lambda_{\gamma' \nu}^\eta \notag \\
    &= \dfrac{g}{3}D(\tau)\sum_{\eta,\gamma,\gamma'} \lambda^\eta_{\mu\gamma} \gloc_{\gamma\sigma,\gamma'\sigma'}(\tau) \lambda_{\gamma' \nu}^\eta, \label{eq:sigma_ep_matrix}
\end{align}

which in matrix form is written

\begin{align}
    \begin{split}
    \hat{\Sigma}^{ep}(\tau) &= \frac{g}{3}D(\tau)\sum_{\eta} \left[\mathsf{g}_{d}(\tau)\hat{\lambda}^\eta \hat{\lambda}^\eta \right. \\ &\phantom{=} \hspace{36pt}\left. + \mathsf{g}_{od}(\tau)\hat{\lambda}^\eta \hat{\mathcal{V}}\hat{\lambda}^\eta \right]  \\
    &= \frac{g}{3}D(\tau)\sum_{\eta} \left[\frac{4}{3} \mathsf{g}_{d}(\tau) \hat{\mathbb{I}}  - \frac{2}{3}\mathsf{g}_{od}(\tau) \hat{\mathcal{V}}\right]. 
    \end{split} \label{eq:sigma_ep_simplified}
\end{align}
where we use the properties of Gell-Mann matrices of indices $\eta = 3, 8$ to obtain the final result. The expression of Equation \ref{eq:sigma_ep_simplified} demonstrates that the electron-phonon self-energy has the structure assumed in Eq. \ref{eqn_diagforsigma}.
\\
\subsection{Propagators and self-energies for electron-electron interactions: Hartree-Fock contributions}

We approximate electron-electron interactions using first-order diagrams, representing the mean-field potential within the Hartree-Fock approximation, and second-order diagrams treated within the second Born approximation. These calculations assume the locality of Coulomb interactions, neglecting Coulombic correlations between electrons at different sites. The Hartree-Fock approximation is time-local, meaning it is proportional to $\delta(\tau)$. In the orbital basis, it is expressed as

\begin{widetext}
\begin{equation} \label{eqn_SigmaHF}
\begin{aligned}
\Sigma^{HF}_{\alpha\sigma,\alpha'\sigma'} &= 
-\sum_{\gamma,\gamma',\sigma_1,\sigma_2} 
\gloc_{\gamma\sigma_1,\gamma'\sigma_2}(\beta) 
\left[
\langle \alpha\gamma'||\alpha'\gamma \rangle \delta_{\sigma\sigma'}\delta_{\sigma_1 \sigma_2} 
- \langle \alpha\gamma'||\gamma\alpha' \rangle \delta_{\sigma\sigma_1} \delta_{\sigma'\sigma_2} 
\right] \\
&= -\left[ 
\delta_{\sigma\sigma'} \sum_{\gamma,\gamma',\sigma_1}  
\gloc_{\gamma\sigma_1,\gamma'\sigma_1}(\beta) \langle \alpha\gamma'||\alpha'\gamma\rangle 
\right] 
+ \sum_{\gamma,\gamma'} 
\gloc_{\gamma\sigma,\gamma'\sigma'}(\beta) \langle \alpha\gamma'||\gamma\alpha' \rangle
\end{aligned}
\end{equation}
\end{widetext}
In this equation, $\beta = \frac{1}{k_B T}$ represents the inverse temperature, and $ \gloc_{\gamma\sigma_1,\gamma'\sigma_1}(\beta)$ denote the matrix elements of the local propagator (Equation \ref{eq:particle_number}). The first term arises from the Hartree diagram, while the second term corresponds to the exchange (Fock) diagram. For orbitals $\alpha_1, \alpha_2, \alpha_3, \alpha_4$ the bare interaction matrix elements are given by

\begin{equation}
\begin{split}
\langle\alpha_1\alpha_2||\alpha_3\alpha_4\rangle &= 
\int \! \int d\vec{r} \, d\vec{r}' \, \psi_{\alpha_1}^*(\vec{r}) \psi_{\alpha_2}(\vec{r}') 
\dfrac{1}{|\vec{r}-\vec{r}'|} \\
&\quad \times \psi_{\alpha_3}(\vec{r}) \psi_{\alpha_4}(\vec{r}')
.
\end{split}
\end{equation}

These Coulomb integrals possess spherical symmetry, which are preserved under the symmetries of the $O_h$ group. Due to this symmetry, we can describe the electron-electron interactions using three independent integrals
\begin{subequations} \label{eq:coulomb}
   \begin{align}
	\langle \alpha\alpha||\alpha\alpha\rangle &= U \\
	\langle \alpha\gamma||\alpha\gamma\rangle &= U'=U-2J \\
	\langle \alpha\gamma||\gamma\alpha\rangle &= J
\end{align} 
\end{subequations}
Now, we can use Eq. \ref{eqn_diag} to derive the Hartree-Fock self-energy terms that are diagonal and off-diagonal, so that $\hat\Sigma^{HF}(\beta) = \mathfrak{s}_{d}^{HF}(\beta)\hat{\mathbb{I}} + \mathfrak{s}_{od}^{HF}(\beta)\hat{\mathcal{V}}$. For the diagonal term, considering that we have $\gamma=\gamma'$ and $\sigma_1 = \sigma_2$, we obtain

\begin{equation} \label{eq:sd_hf}
\begin{split}
\mathfrak{s}_{d}^{HF}(\beta) &= -\sum_\gamma \mathsf{g}_{d}(\beta) 
\Big[ 2\langle \alpha\gamma || \alpha\gamma \rangle 
- \langle \alpha\gamma || \gamma\alpha \rangle \Big] \\
&= -2 \mathsf{g}_{d}(\beta) \left(U + 2U' - J\right)
\end{split}
\end{equation}
while for the off-diagonal term it can be demonstrated that 
\begin{align} 
	\mathfrak{s}_{d}^{HF}(\beta) &= U'\mathsf{g}_{od}(\beta) \label{eq:sd_hf_offdiag}
\end{align}
Equations~\ref{eq:sd_hf} and \ref{eq:sd_hf_offdiag} confirm that the HF electron self-energy retains the structural form assumed in Eq.~\ref{eqn_diagforsigma}.\\

\subsection{Propagators and self-energies for electron-electron interactions: second order Born approximation}

In the second Born approximation there are two kind of topologically non-equivalent diagrams
\begin{widetext}
\begin{subequations} \label{eqn:Sigma2}
\begin{align}
\Sigma^{(2a)}_{\mu\sigma,\nu\sigma'}(\tau) &= 
-\sum \gloc_{\lambda\sigma,\kappa\sigma'}(\tau) \langle \mu\chi||\lambda\alpha \rangle 
\gloc_{\alpha\sigma_1,\beta\sigma_2}(\tau) 
\gloc_{\xi\sigma_2,\chi\sigma_1}(\beta-\tau) 
\langle \beta\kappa||\xi\nu \rangle, \\
\Sigma^{(2b)}_{\mu\sigma,\nu\sigma'}(\tau) &= 
\sum \gloc_{\alpha\sigma_1,\kappa\sigma'}(\tau) \langle \mu\chi || \lambda\alpha \rangle 
\gloc_{\lambda\sigma,\beta\sigma_2}(\tau) 
\gloc_{\xi\sigma_2,\chi\sigma_1}(\beta-\tau) 
\langle \beta\kappa || \xi\nu \rangle.
\end{align}
\end{subequations}
\end{widetext}
Expanding both diagrammatic terms using the Coulomb integrals described in Equations \ref{eq:coulomb} and considering the algebraic structure of the local Green’s function matrix $\hat{\mathcal{G}}$, the two contributions can be combined as $\hat{\Sigma}^B(\beta) = \hat{\Sigma}^{(2a)}(\beta) + \hat{\Sigma}^{(2b)}(\beta) = \mathfrak{s}_{d}^{B}(\beta)\hat{\mathbb{I}} + \mathfrak{s}_{od}^{B}(\beta)\hat{\mathcal{V}}$. This formulation yields both the diagonal and off-diagonal elements of the second Born self-energy
\begin{widetext}
\begin{subequations} \label{eqn:selfE_B}
\begin{align}
\begin{split}
\mathfrak{s}_{d}^B(\tau) &= 
- \left( 5U^2 - 20UJ + 28J^2 \right) \mathsf{g}_{d}^2(\tau)\mathsf{g}_{d}(\beta-\tau) 
- 8\left( U^2 - 4UJ + 3J^2 \right) \mathsf{g}_{d}(\tau)\mathsf{g}_{od}(\tau)\mathsf{g}_{od}(\beta-\tau) \\
&\quad + 2\left( U^2 - 4UJ + 5J^2 \right) \mathsf{g}_{od}^2(\tau)\left[ \mathsf{g}_{d}(\beta-\tau) - \mathsf{g}_{od}(\beta-\tau) \right],
\end{split} \\
\begin{split}
\mathfrak{s}_{od}^B(\tau) &= 
\left( U^2 - 4UJ + 5J^2 \right) \mathsf{g}_{d}^2(\tau)\mathsf{g}_{od}(\beta-\tau)
- 2\left( U^2 - 2UJ + 3J^2 \right) \mathsf{g}_{d}(\tau)\mathsf{g}_{od}(\tau)\left[ 2\mathsf{g}_{d}(\beta-\tau) + \mathsf{g}_{od}(\beta-\tau) \right] \\
&\quad - \left( U^2 - 4UJ + 3J^2 \right) \mathsf{g}_{od}^2(\tau)\mathsf{g}_{d}(\beta-\tau)
- \left( 9U^2 - 36UJ + 38J^2 \right)\mathsf{g}_{od}^2(\tau)\mathsf{g}_{od}(\beta-\tau).
\end{split}
\end{align}
\end{subequations}
\end{widetext}

These expressions confirm that the second order Born approximation electron self-energy retains the structural form assumed in Eq.~\ref{eqn_diagforsigma}.

\subsection{Propagators and self-energies for electron-electron interactions: orbital exchange}

The orbital exchange term is

\begin{align}
    \hamil_{oex} &= \sum \mathcal J_{ij}^{\eta\eta'} c_{i\alpha\sigma}^\dagger \lambda^\eta_{\alpha\beta} c_{i\beta\sigma} c_{j\mu\sigma'}^\dagger \lambda^{\eta'}_{\mu\nu} c_{j\nu\sigma'}.
\end{align}

This interaction must respect the symmetry operations of the $O_h$ group. First, note that time-reversal symmetry enforces $\mathcal J^{\eta\eta'}_{ij} = \mathcal J^{\eta'\eta}_{ij}$. From this, we can deduce that the coupling matrix elements satisfy $\mathcal J^{\eta\eta'} = \delta_{\eta\eta'} \mathcal J^{\eta\eta'}$. Furthermore, the subsets of Gell-Mann matrices $\{\lambda^1, \lambda^4, \lambda^6\}$, $\{\lambda^2, \lambda^5, \lambda^7\}$, and $\{\lambda^3, \lambda^8\}$ form three distinct irreducible representations of the $O_h$ group, corresponding to the $T_{2g}$, $T_{1g}$, and $E_g$ representations, respectively. Consequently, the coupling coefficients must be equal within each subset.\\

The Jahn-Teller distortions we consider belong to the $E_g$ irreducible representation, whose associated Gell-Mann matrices are diagonal. As a result, we can simplify the Hamiltonian to
\begin{align}
\hamil_{oex} &= - \sum_{i,j} \mathcal{J}^{'} \sum_{\eta \in \{3,8\}} \sum_{\alpha,\sigma} \sum_{\alpha',\sigma'} 
c_{i\alpha\sigma}^\dagger \lambda^\eta_{\alpha\alpha} c_{i\alpha\sigma} \\
&\quad \times c_{j\alpha'\sigma'}^\dagger \lambda^\eta_{\alpha'\alpha'} c_{j\alpha'\sigma'}.
\end{align}

We employ a Hartree-Fock approximation to construct the orbital exchange self-energy:
\begin{align}
    \Sigma^{oex}_{\alpha\beta} &= -\sum_{\mu,\nu} G_{\mu\nu}(\beta) \left[ \langle \alpha\nu||\mu\beta \rangle - \langle \alpha\nu||\beta\mu \rangle \right]
    \label{eq:sigma_oex}
\end{align}
From the definition of the matrix elements, we infer that the only nonzero terms for the exchange interactions between sites $i$ and $j$ are given by
\begin{align}
    \langle i\alpha\sigma, j\alpha'\sigma' || i\alpha\sigma, j\alpha'\sigma' \rangle = -\mathcal J^{'} \sum_\eta \Lambda_{\alpha\alpha'}
\end{align}
where we define $\Lambda_{\alpha\alpha'} = \sum_\eta \lambda^\eta_{\alpha\alpha} \lambda^\eta_{\alpha'\alpha'}$. This allows us to analyze the two terms that arise from Equation \ref{eq:sigma_oex} (i.e., Hartree and Fock) separately

\begin{subequations}
    \begin{align}
        \Sigma^{eox, H}_{i\alpha\sigma,j\alpha'\sigma'} &= \mathcal J^{'} \Lambda_{\alpha\alpha'} G_{i\alpha\sigma,j\alpha'\sigma'}(\beta) \\
        \Sigma^{eox, F}_{i\alpha\sigma,i\alpha\sigma} &= -\mathcal J^{'} \sum_{j,\alpha',\sigma'} \Lambda_{\alpha\alpha'} G_{j\alpha'\sigma',j\alpha'\sigma'}(\beta)
    \end{align}
\end{subequations}

In momentum space, considering that the orbital exchange interaction is restricted to nearest neighbors, we obtain

\begin{widetext}
\begin{subequations} \label{eqn:eox_HF}
\begin{align}
\Sigma^{eox, H}_{\vec{k}\alpha\sigma,\vec{k}\alpha'\sigma'} &= 
\frac{1}{\mathcal{N}^2} \sum_{i,j} e^{\imath(\vec{k}-\vec{q})\cdot\vec{r}_{ij}} 
\mathcal{J}^{'} \Lambda_{\alpha\alpha'} G_{\vec{q}\alpha\sigma,\vec{q}\alpha'\sigma'}(\beta) 
= \frac{1}{\mathcal{N}} \sum_{\vec{q}} \gamma_{\vec{k}-\vec{q}} \Lambda_{\alpha\alpha'} 
G_{\vec{q}\alpha\sigma,\vec{q}\alpha'\sigma'}(\beta), \\
\Sigma^{eox, F}_{\vec{k}\alpha\sigma,\vec{k}\alpha\sigma} &= 
- \frac{1}{\mathcal{N}^2} \sum_{i,j} \sum_{\alpha',\vec{q}} \mathcal{J}^{'} \Lambda_{\alpha\alpha'} 
G_{\vec{q}\alpha'\sigma',\vec{q}\alpha'\sigma'}(\beta) 
= -\gamma_{\vec{0}} \sum_{\alpha'} \Lambda_{\alpha\alpha'} 
\mathcal{G}_{\alpha'\sigma',\alpha'\sigma'}(\beta).
\end{align}
\end{subequations}
\end{widetext}

Here, we define

\begin{align}
    \gamma_{\vec{k}} &= \sum_j e^{\imath\vec{k} \cdot \vec{r}_{ij}} \mathcal J^{'}
\end{align}

For a cubic lattice, this simplifies to

\begin{align}
    \gamma_{\vec{k}} &= 2\mathcal J^{'} \sum_{w \in \{x,y,z\}} \cos(k_w a)
\end{align}

Finally, to verify whether the assumed symmetry is preserved, we inspect the matrix elements of $\hat{\Lambda}$

\begin{align}
    \Lambda_{\alpha\alpha'} &= 2\delta_{\alpha\alpha'} - \frac{2}{3}
\end{align}

This results in the following diagonal and off-diagonal terms of the orbital exchange matrix $\hat{\Sigma}^{oex}(\beta) = \mathfrak{s}_{d, \vec{k}}^{oex}(\beta)\hat{\mathbb{I}} + \mathfrak{s}_{od, \vec{k}}^{oex}(\beta)\hat{\mathcal{V}}$, expressed as

\begin{subequations} \label{eqn:oex_selfenergy}
\begin{align}
    \mathfrak{s}_{d, \vec{k}}^{oex}(\beta) &= \frac{4}{3\mathcal{N}} \sum_{\vec{q}} \gamma_{\vec{k}-\vec{q}} \mathrm{g}_{d, \vec{q}}(\beta) \label{eqn:oex_diag} \\
    \mathfrak{s}_{od, \vec{k}}^{oex}(\beta) &= -\frac{2}{3\mathcal{N}} \sum_{\vec{q}} \gamma_{\vec{k}-\vec{q}} \mathrm{g}_{od, \vec{q}}(\beta) \label{eqn:oex_offdiag}
\end{align}
\end{subequations}
where we recall that $\mathcal{N}$ is the number of grid points into reciprocal space. Equations \ref{eqn:oex_selfenergy} confirm that the structure of the HF self-energy remains consistent with the form assumed in Eq.~\ref{eqn_diagforsigma}.

\section{Orbital and spin-orbital correlations}\label{app_corr}
\renewcommand{\theequation}{E\arabic{equation}}

To investigate spin, orbital and spin-orbital fluctuations, we focus on two-particle correlations, described by the following Green’s function in the imaginary-time axis
\begin{align}
    G^{(2)}_{\alpha_1\alpha_2\alpha_3\alpha_4}(\tau_1,\tau_2,\tau_3,\tau_4) &= \left\langle \mathcal{T} c_{\alpha_1}^\dagger(\tau_1) c_{\alpha_2}(\tau_2) c_{\alpha_3}^\dagger(\tau_3) c_{\alpha_4}(\tau_4) \right\rangle.
\end{align}
which can be expressed as
\begin{align}
\begin{split}
G^{(2)}_{\alpha_1\alpha_2\alpha_3\alpha_4}(\tau_1,\tau_2,\tau_3,\tau_4) &= 
G_{\alpha_2\alpha_1}(\tau_2{-}\tau_1) 
G_{\alpha_4\alpha_3}(\tau_4{-}\tau_3) \\
&\quad - G_{\alpha_2\alpha_3}(\tau_2{-}\tau_3) 
G_{\alpha_4\alpha_1}(\tau_4{-}\tau_1) \\
&\quad + \Upsilon_{\alpha_1\alpha_2\alpha_3\alpha_4}(\tau_1,\tau_2,\tau_3,\tau_4).
\end{split}
\end{align}

The term \( \Upsilon_{\alpha_1\alpha_2\alpha_3\alpha_4}(\tau_1,\tau_2,\tau_3,\tau_4) \) accounts for corrections beyond the mean-field approximation. However, in this paper, we focus on mean-field correlations, and thus we approximate \( \Upsilon \approx 0 \)

\begin{align}
\begin{split}
G^{(2)}_{\alpha_1\alpha_2\alpha_3\alpha_4}(\tau_1,\tau_2,\tau_3,\tau_4) &\approx 
G_{\alpha_2\alpha_1}(\tau_2{-}\tau_1)G_{\alpha_4\alpha_3}(\tau_4{-}\tau_3) \\
&\quad - G_{\alpha_2\alpha_3}(\tau_2{-}\tau_3)G_{\alpha_4\alpha_1}(\tau_4{-}\tau_1).
\end{split}
\end{align}

To give a general framework that can be applied to all sorts of correlations, let us define define two generic local operators as
\begin{align}
    U &= \sum_{i,\alpha,\beta} u_{\alpha\beta} c_{i\alpha}^\dagger c_{i\beta}, \\
    V &= \sum_{i,\alpha,\beta} v_{\alpha\beta} c_{i\alpha}^\dagger c_{i\beta},
\end{align}
and compute their correlations on different lattice sites $i,j$
\begin{align}
    \left\langle U_i V_j \right\rangle &= \sum_{\alpha,\beta} \sum_{\alpha',\beta'} u_{\alpha\beta} v_{\alpha'\beta'} \left\langle c_{i\alpha}^\dagger c_{i\beta} c_{j\alpha'}^\dagger c_{j\beta'} \right\rangle.
\end{align}
This relates to \( G^{(2)} \) in the ordered time limit \( \tau_1 > \tau_2 > \tau_3 > \tau_4 \to 0 \):
\begin{align} \label{eq:correlUV}
\left\langle U_i V_j \right\rangle &\approx 
\sum_{\alpha,\beta} u_{\alpha\beta} \gloc_{\beta\alpha}(\beta) 
\sum_{\alpha',\beta'} v_{\alpha'\beta'} \gloc_{\beta'\alpha'}(\beta) \\
&\quad + \sum_{\alpha,\beta} \sum_{\alpha',\beta'} 
u_{\alpha\beta} v_{\alpha'\beta'} G_{i\beta,j\alpha'}(0) G_{j\beta',i\alpha}(\beta).
\end{align}

The covariance is given by
\begin{align}
    \left\langle \delta U_i \delta V_j \right\rangle &= \left\langle U_i V_j \right\rangle - \left\langle U_i  \right\rangle\left\langle V_j \right\rangle,
\end{align}
and the first term of Equation \ref{eq:correlUV} is the product of expected values, so the covariance on the real space can be expressed as
\begin{align}
    \left\langle \delta U_i \delta V_j \right\rangle &\approx \sum_{\alpha,\beta} \sum_{\alpha',\beta'} u_{\alpha\beta} v_{\alpha'\beta'} G_{i\beta,j\alpha'}(0)G_{j\beta',i\alpha}(\beta).
\end{align}

Since Green's functions are diagonal in $\vec{k}$-space, it is convenient to perform the spatial Fourier transform

\begin{align}
\left\langle \delta U_i \delta V_j \right\rangle &\approx 
\dfrac{1}{\mathcal{N}^2} \sum_{\alpha,\beta} \sum_{\alpha',\beta'} 
\sum_{\vec{k},\vec{k}'} u_{\alpha\beta} v_{\alpha'\beta'} \ 
e^{-i(\vec{k}-\vec{k}')\cdot \vec{r}_{ij}} \\
&\quad \times G_{\beta\alpha';\vec{k}}(0) G_{\beta'\alpha;\vec{k}'}(\beta).
\end{align}

By introducing the momentum transfer \( \vec{q} = \vec{k} - \vec{k}' \), we observe that the mean-field covariance also becomes diagonal in reciprocal space

\begin{align} \label{eq:general_covar}
    \left\langle \delta U_\vec{q} \delta V_\vec{-q} \right\rangle &\approx \frac{1}{\mathcal{N}} \sum_\vec{k} \ \text{Tr} \left[ \hat{u} \hat{G}_\vec{k}(0) \hat{v} \hat{G}_{\vec{k}-\vec{q}}(\beta) \right].
\end{align}

Focusing on orbital correlations, we note that the detection of the Jahn-Teller modes can be done by orbital-charge operators
\begin{align}
   \mathbb{T}_{i\eta} &= \sum_{\alpha,\alpha',\sigma} \lambda_{\alpha\alpha'}^\eta c_{i\alpha\sigma}^\dagger c_{i\alpha'\sigma}.
\end{align}
The two Gell-Mann matrices considered are those corresponding to tetragonal and orthorhombic modes of distortion, $\hat{\lambda}^8$ and $\hat{\lambda}^3$, respectively. Those matrices only reflect distortions along z-axis, so we can make linear combinations
\begin{subequations}
    \begin{align}
        \hat{\lambda}^8(n) &= \cos\dfrac{2n\pi}{3} \hat{\lambda}^8 - \sin\dfrac{2n\pi}{3} \hat{\lambda}^3 \\
        \hat{\lambda}^3(n) &= \sin\dfrac{2n\pi}{3} \hat{\lambda}^8 + \cos\dfrac{2n\pi}{3} \hat{\lambda}^3
    \end{align}
\end{subequations}

With \( n = 0,1,2 \), we analyze distortions along the other Cartesian axes by correlating the modified orbital-charge operators

\begin{align}
    \mathbb{T}_{i\eta}(n) &= \sum_{\alpha,\alpha',\sigma} \lambda_{\alpha\alpha'}^\eta(n) c_{i\alpha\sigma}^\dagger c_{i\alpha'\sigma}.
\end{align}

Developing the trace of Equation \ref{eq:general_covar} we get

\begin{widetext}
\begin{subequations} \label{eq:T_correlations}
\begin{align}
\left\langle \delta \mathbb{T}_{\theta,\vec{q}}(m) \delta \mathbb{T}_{\theta,\vec{-q}}(n) \right\rangle &= 
\dfrac{4}{\mathcal{N}} \cos\left( \dfrac{2(m{-}n)\pi}{3} \right) \sum_\vec{k} 
\left[ \mathrm{g}_{d, \vec{k}}(0)\, \mathrm{g}_{d, \vec{k}-\vec{q}}(\beta) 
- \mathrm{g}_{od, \vec{k}}(0)\, \mathrm{g}_{od, \vec{k}-\vec{q}}(\beta) \right], \\
\left\langle \delta \mathbb{T}_{\phi,\vec{q}}(m) \delta \mathbb{T}_{\phi,\vec{-q}}(n) \right\rangle &= 
\dfrac{4}{\mathcal{N}} \cos\left( \dfrac{2(m{-}n)\pi}{3} \right) \sum_\vec{k} 
\left[ \mathrm{g}_{d, \vec{k}}(0)\, \mathrm{g}_{d, \vec{k}-\vec{q}}(\beta) 
- \mathrm{g}_{od, \vec{k}}(0)\, \mathrm{g}_{od, \vec{k}-\vec{q}}(\beta) \right], \\
\left\langle \delta \mathbb{T}_{\theta,\vec{q}}(m) \delta \mathbb{T}_{\phi,\vec{-q}}(n) \right\rangle &= 
\dfrac{4}{\mathcal{N}} \sin\left( \dfrac{2(m{-}n)\pi}{3} \right) \sum_\vec{k} 
\left[ \mathrm{g}_{d, \vec{k}}(0)\, \mathrm{g}_{d, \vec{k}-\vec{q}}(\beta) 
- \mathrm{g}_{od, \vec{k}}(0)\, \mathrm{g}_{od, \vec{k}-\vec{q}}(\beta) \right].
\end{align}
\end{subequations}
\end{widetext}

To analyze the cooperative nature of orbital correlations, we compute the expectation values \( \langle \delta \mathbb{T}_{\eta, \vec{q}}(m) \delta \mathbb{T}_{\zeta, \vec{-q}}(n) \rangle \), where \( \eta, \zeta \) correspond to tetragonal (\(\theta\)) and orthorhombic (\(\phi\)) Jahn-Teller distortions, and \( m, n = 0,1,2 \) label spatial orientations along Cartesian directions. Importantly, the structure of the correlation functions in Equations \ref{eq:T_correlations} provides direct insight into the nature of orbital ordering as we discuss in the following. \\

In this regard, the presence of the cosine and sine terms, \( \cos \frac{2(m - n) \pi}{3} \) and \( \sin \frac{2(m - n) \pi}{3} \), in Equations \ref{eq:T_correlations} reflects how distortions along different axes interact, allowing for a detailed characterization of orbital correlations. A key feature of these correlations is their ability to distinguish between ferro-orbital order, where orbitals align uniformly across sites, and cooperative Jahn-Teller (JT) distortions, where orbitals alternate directions in a staggered pattern. Specifically, for \( m = n \), corresponding to ferro-orbital correlations where JT modes at two given sites are aligned in the same direction, positive correlations should be observed:  
\[
\left\langle \delta  \mathbb{T}_{\theta,\vec{q}}(m) \delta \mathbb{T}_{\theta,\vec{-q}}(m) \right\rangle, \left\langle \delta \mathbb{T}_{\phi,\vec{q}}(m) \delta \mathbb{T}_{\phi,\vec{-q}}(m) \right\rangle > 0.
\]  
Conversely, cooperative orbital order, where distortions at different sites are arranged orthogonally (\( m \neq n \)), is characterized by negative correlations  
\[
\left\langle \delta \mathbb{T}_{\theta,\vec{q}}(m) \delta  \mathbb{T}_{\theta,\vec{-q}}(n) \right\rangle, \left\langle \delta \mathbb{T}_{\phi,\vec{q}}(m) \delta \mathbb{T}_{\phi,\vec{-q}}(n) \right\rangle < 0.
\]  
This orthogonal arrangement of distortions is a hallmark of cooperative JT order, where orbital distortions propagate in an alternating pattern across the lattice. This distinction is particularly evident in the computed correlations of the same JT mode, \( \langle \delta \mathbb{T}_{\theta,\vec{q}}(m) \delta \mathbb{T}_{\theta,\vec{-q}}(n) \rangle \) and \( \langle \delta \mathbb{T}_{\phi,\vec{q}}(m) \delta \mathbb{T}_{\phi,\vec{-q}}(n) \rangle \), which determine whether tetragonal or orthorhombic distortions evolve uniformly or staggered throughout the system. Additionally, the cross-correlation \( \langle \delta \mathbb{T}_{\theta,\vec{q}}(m) \delta \mathbb{T}_{\phi,\vec{-q}}(n) \rangle \) quantifies the coupling between tetragonal and orthorhombic distortions, revealing whether these two modes coexist or compete within the system.\\  

Beyond purely orbital correlations, we further explore the interplay between spin and orbital degrees of freedom by introducing the spin-charge operator
\begin{align}
     \mathbb{W} &= \sum_{i} \sum_{\eta,\alpha,\alpha'} \sum_{k,\sigma,\sigma'} \lambda^\eta_{\alpha\alpha'} \sigma^k_{\sigma\sigma'} c_{i\alpha\sigma}^\dagger c_{i\alpha'\sigma'}.
\end{align}  
This operator captures the coupling between local orbital configurations and spin orientations, allowing us to assess how distortions influence spin dynamics. The covariance in reciprocal space is given by

\begin{align}
\left\langle \delta \mathbb{W}_\vec{q} \delta \mathbb{W}_\vec{-q} \right\rangle &= 
\frac{2}{\mathcal{N}} \sum_\vec{k} \big[ 
3\,\mathrm{g}_{d, \vec{k}}(0)\mathrm{g}_{d, \vec{k}-\vec{q}}(\beta) \\
&\quad + \mathrm{g}_{od, \vec{k}}(0)\mathrm{g}_{od, \vec{k}-\vec{q}}(\beta) \big].
\end{align}
which provides direct insight into spin-orbital correlations across the lattice. The structure of this expression suggests that spin-orbital interactions are mediated by both diagonal (\( \mathrm{g}_{d, \vec{k}} \)) and off-diagonal (\( \mathrm{g}_{od, \vec{k}} \)) Green’s functions, indicating a nontrivial mixing of spin and orbital excitations. Notably, the prefactor of 3 in front of \( \mathrm{g}_{d, \vec{k}} \) highlights the dominant contribution of diagonal orbital terms to spin-orbital fluctuations.

These correlations play a crucial role in understanding emergent magnetic and electronic behaviors in strongly correlated systems, as spin-orbital entanglement can drive novel phases such as spin-orbital liquids or magnetically ordered states with intertwined orbital dynamics. Our analysis thus provides a comprehensive framework for investigating the interplay between lattice distortions, orbital ordering, and spin correlations, offering deeper insights into the fundamental mechanisms governing spin-orbital excitations.

\section{Computation of real-space correlators}
\renewcommand{\theequation}{F\arabic{equation}}

The real-space correlators \( \langle \delta \mathbb{T}_i \delta \mathbb{T}_j \rangle \) and \( \langle \delta \mathbb{W}_i \delta \mathbb{W}_j \rangle \) were computed from their corresponding momentum-space representations via an inverse Fourier transform. To achieve this, we employed a fine discretization of the Brillouin zone (BZ), using a dense grid composed of \( 24 \times 24 \times 24 \) k-points.  The calculations were initially carried out within the irreducible Brillouin zone (IBZ), significantly reducing the computational effort. By exploiting the inherent octahedral symmetry \( O_h \) of the crystal lattice, we extended the computed results from the IBZ to fill the full BZ. This symmetry operation exploits the invariance of the system under rotations and reflections consistent with the octahedral point group, effectively reproducing the full BZ from the smaller IBZ dataset without sacrificing accuracy. Importantly, the weights of each point of the IBZ to be extrapolated to the whole BZ are intrinsically taken into account by considering all possible \( O_h \) point symmetry operations.\\

The inverse Fourier transform that relates the real-space correlators for two general operators, $U,V$, to their momentum-space counterparts is given explicitly by the equation
\begin{equation}
\langle \delta U_i \delta V_j \rangle = \frac{1}{\mathcal{N}} \sum_{\vec{q} \in \text{BZ}} \langle \delta U_{\vec{q}} \delta V_{\vec{-q}} \rangle e^{-i\vec{q} \cdot \vec{r}_{ij}}.
\end{equation}
Here, \( \mathcal{N} \) represents the total number of k-points sampled in the full BZ, ensuring proper normalization. Each term in this summation captures contributions from different momentum-space fluctuations modulated by the complex exponential factor \( e^{-i\vec{q}\cdot\vec{r}} \), which encodes spatial phase relationships crucial for understanding the spatial distribution and coherence length of orbital and spin fluctuations in the material.\\

Correlators for the operators $\mathbb{T}$ and $\mathbb{W}$ can be transformed in this manner, providing a spatially resolved picture of fluctuations and revealing how local perturbations propagate through the lattice. By analyzing these real-space correlators, one can infer the emergence of short-range order phenomena driven by interactions such as spin-orbit coupling and Jahn-Teller distortions. These effects become particularly pronounced when band hybridization is incorporated through the inclusion of hopping amplitudes.

\section{Energy and Galitzkii-Migdal formula} \label{Galitzkii}
\renewcommand{\theequation}{G\arabic{equation}}

In quantum theory the expected value of an observable $\hat{\mathcal{O}}$ is taken from the trace of its product with density matrix $\hat{\rho}$
\begin{align}
    \left\langle \hat{\mathcal{O}} \right\rangle &= \operatorname{Tr}\left[ \hat{\rho} \hat{\mathcal{O}} \right]
    .
\end{align}
The density matrix is defined through creation and annihilation operators
\begin{align}
    \rho_{\mu\nu} &= \left\langle \hat{c}_\nu^\dagger \hat{c}_\mu \right\rangle
\end{align}
which is similar to the definition of the quantum Green's function in the imaginary time
\begin{align}
    G_{\mu\nu}(\tau) &= -\left\langle \hat{\mathcal{T}} \hat{c}_\mu(\tau) \hat{c}_\nu^\dagger \right\rangle.
\end{align}

The connection to the density matrix emerges in the limit \( \tau \to 0^- \), where the Green’s function retains information about the system’s quantum state occupancy and correlations

\begin{align}
    \lim_{\tau \to 0^-} G_{\mu\nu}(\tau) &= \mp\lim_{\tau\to 0^-} \left\langle \hat{c}_\nu^\dagger \hat{c}_\mu(\tau) \right\rangle = \mp \rho_{\mu\nu}
\end{align}
and using the periodicity property of Matsubara Green's function $\hat{G}(\beta-\tau) = \pm \hat{G}(\tau)$
\begin{align}
    \rho_{\mu\nu} &= -G_{\mu\nu}(\beta)
    .
\end{align}

We compute the energy contributions from all terms in the Hamiltonian. To achieve this, we select the appropriate Green’s function to evaluate the trace of the quadratic terms. Since we are working with an infinite lattice, we consider the local energy density $\mathcal{E} = E/\mathcal{N}$, where $\mathcal{N}$ is the number of sites. The simplest case corresponds to local quadratic terms, such as spin-orbit coupling

\begin{align}
\mathcal{E}_{\text{pot}} &= \frac{1}{\mathcal{N}} \sum_{i, \alpha, \beta, \sigma, \sigma'} 
    \mathcal{V}_{\alpha\sigma, \beta\sigma'} \, \rho_{i\beta\sigma', i\alpha\sigma}(t) = \operatorname{Tr} \left[\hat{\gloc}(\beta) \hat{\mathcal{V}} \right].
\end{align}

Similarly, the kinetic energy is given by

\begin{align}
\mathcal{E}_{\text{kin}} &= -\frac{1}{\mathcal{N}} 
\sum_{i,j, \alpha, \beta, \sigma} 
t_{i\alpha, j\beta} \, \rho_{j\beta\sigma, i\alpha\sigma} \notag \\
&= \frac{1}{\mathcal{N}} 
\sum_{\vec{k}, \alpha, \beta, \sigma} 
\varepsilon_{\alpha\beta;\vec{k}} \, 
\rho_{\beta\sigma, \alpha\sigma; \vec{k}} \notag \\
&= -\frac{1}{\mathcal{N}} 
\sum_{\vec{k}} \operatorname{Tr} 
\left[\hat{\gloc}_\vec{k}(\beta) \hat{\varepsilon}_\vec{k} \right].
\end{align}

For the electron-electron interaction term, we separate the Hamiltonian into $\hat{\hamil} = \hat{\hamil}_0 + \hat{\hamil}_{\text{int}}$, where $\hat{\hamil}_0$ is the quadratic part, expressed as a sum over $\varepsilon_{\alpha\beta} c_\alpha^\dagger c_\beta$, and $\hat{\hamil}_{\text{int}}$ includes interaction terms involving more than two creation-annihilation operators, such as $u_{\alpha\beta}^\phi c_\alpha^\dagger c_\beta \phi$, where $\phi$ represents any combination of fermionic and bosonic operators.\\

Recalling the definition of the Green’s function

\begin{align}
G_{\alpha\beta}(\tau) &= - \left\langle \mathcal{T} 
    c_\alpha(\tau) \, c_\beta^\dagger \right\rangle,
\end{align}

and using the equation of motion for the annihilation operator,

\begin{align}
\partial_\tau c_\alpha(t) &= [\hat{\hamil}, c_\alpha](\tau),
\end{align}

along with the fermionic commutation relation,

\begin{align}
[c_\mu^\dagger c_\gamma, c_\alpha] &= -\delta_{\alpha\mu} c_\gamma,
\end{align}

and the time-ordering property $\Theta(\tau)$, we derive the equation for the Green’s function

\begin{align}
-\partial_\tau G_{\alpha\beta}(\tau) &= \delta(\tau) \delta_{\alpha\beta} 
+ \sum_\gamma \varepsilon_{\alpha\gamma} G_{\gamma\beta}(\tau) \notag \\
&\phantom{=} - \sum_{\mu, \gamma, \phi} u_{\mu\gamma}^\phi \, 
\left\langle \mathcal{T} \left[ c_\mu^\dagger c_\gamma \phi,\, c_\alpha \right](\tau) 
c_\beta^\dagger \right\rangle.
\end{align}

By analogy with the Dyson equation, we relate the last term to the self-energy

\begin{align}
	\left[\hat{G} * \hat{\Sigma}\right]_{\alpha\beta}(\tau) &= - \sum_{\mu, \gamma, \phi} u_{\mu\gamma}^\phi \left\langle \mathcal{T} \left[ c_\mu^\dagger c_\gamma \phi, c_\alpha \right](\tau) c_\beta^\dagger \right\rangle.
\end{align}

To evaluate the commutator, we compute

\begin{align}
	\left[ c_\mu^\dagger c_\gamma \phi, c_\alpha \right] &= c_\mu^\dagger c_\gamma \phi c_\alpha - c_\alpha c_\mu^\dagger c_\gamma \phi \\
    &= \left[ c_\mu^\dagger c_\gamma, c_\alpha \right] \phi + c_\mu^\dagger c_\gamma \left[ \phi, c_\alpha \right] \\
    &= -\delta_{\alpha\mu} c_\gamma \phi + c_\mu^\dagger c_\gamma \left[ \phi, c_\alpha \right].
\end{align}

For electron-phonon interactions, $\phi$ is a bosonic operator, implying $\left[ \phi, c_\alpha \right] = 0$. However, for electron-electron interactions, $\phi$ consists of a pair of fermionic creation and annihilation operators, introducing an additional Kronecker delta in $\alpha$. Due to the Pauli exclusion principle, only one term survives:

\vspace{0.5em}
\begin{align}
	F_{\alpha\beta}(\tau) =\left[\hat{G} * \hat{\Sigma}\right]_{\alpha\beta}(\tau) 
	= \sum_{\gamma, \phi} u_{\alpha\gamma}^\phi 
	\left\langle \mathcal{T} c_\gamma(\tau) \phi(\tau) c_\beta^\dagger \right\rangle.
\end{align}

Here, “$*$” denotes convolution in imaginary time. Finally, taking the trace in the limit $\tau \to 0^-0$ yields the expectation value of the interaction energy (electron-electron, electron-phonon, etc.):

\begin{align}
	\mathcal{E}_{\text{int}} &= -\operatorname{tr} \left[ \hat{F}(\beta) \right].
\end{align}

\bibliographystyle{apsrev4-2_maxauth} % We choose the "plain" reference style
\bibliography{biblionegf} % Entries are in the refs.bib file

\end{document}